\documentclass[aps,prd,twocolumn,groupedaddress,amsmath,amssymb,floatfix]{revtex4-2}
\usepackage{lineno,graphicx,color,hyperref,bm, orcidlink,placeins, mathtools}
\usepackage{tikz-feynhand, contour,here,multirow}

\newcommand{\citeall}{\cite{benhar2006archivequasielasticelectronnucleusscattering,Barreau:1983ht,Baran:1988tw,Sealock:1989nx,Bagdasaryan:1988hp,Day:1993md,Anghinolfi:1996vm,O'Connell:1987ag}}
\newcommand{\citecarbon}{\cite{benhar2006archivequasielasticelectronnucleusscattering,Barreau:1983ht,Baran:1988tw,Sealock:1989nx,Bagdasaryan:1988hp,Day:1993md}}

\begin{document}
\preprint{APS/123-QED}

%%% Title %%%
\title{Implementation and investigation of electron-nucleus scattering in the \textsc{NEUT} neutrino event generator}

%%% Author list %%%
\newcommand{\kamioka}{\affiliation{Kamioka Observatory, Institute for Cosmic Ray Research,
University of Tokyo, Kamioka, Gifu 506-1205, Japan}}
\author{Seisho Abe\,\orcidlink{0000-0002-2110-5130}} \kamioka
\email{seisho@icrr.u-tokyo.ac.jp}
\date{\today}

%%% Abstract %%%
\begin{abstract}
Understanding nuclear effects is essential for improving the sensitivity of neutrino oscillation measurements.
Validating nuclear models solely through neutrino scattering data is challenging due to limited statistics and the broad energy spectrum of neutrinos.
In contrast, electron scattering experiments provide abundant high-precision data with various monochromatic energies and angles.
Since both neutrinos and electrons interact via electroweak interactions, the same nuclear models can be applied to simulate both interactions.
Thus, high-precision electron scattering data is essential for validating the nuclear models used in neutrino experiments.
To enable this, the author has introduced a new electron scattering framework in the \textsc{NEUT} neutrino event generator, covering two interaction modes: quasielastic (QE) and single pion production.
\textsc{NEUT} predictions of QE agree well with numerical calculations, supporting the validity of this implementation.
From comparisons with \textsc{NEUT} predictions and inclusive electron scattering data, the momentum-dependent removal energy correction is derived, addressing effects beyond the plane wave impulse approximation.
This correction is applied to neutrino interactions, observing significant changes in charged lepton kinematics.
Notably, the reconstructed neutrino energy distribution shows a peak shift of approximately 20--30\,MeV, which is crucial for accurately measuring neutrino oscillation parameters.
\end{abstract}
\maketitle

%%%%%%%%%%%%%%%%%%%%%%%%%%%%%
\section{Introduction} \label{sec:intro}
Neutrino-nucleus interaction is one of the dominant systematic uncertainties in long-baseline neutrino oscillation experiments, such as T2K and NOvA experiments~\cite{Abe2023,PhysRevD.108.072011,PhysRevD.106.032004}.
The neutrino energy is reconstructed using measured particle kinematics.
This analysis relies on neutrino-nucleus interaction modeling, which involves complex multibody physics processes.
Various neutrino experiments measure and constrain the models, but substantial uncertainties remain.
The difficulties mainly arise from two factors.
First, we have limited statistics due to the small neutrino cross section.
Neutrino detectors are typically large to gather sufficient statistics, but coarse measurements around the target nucleus.
Second, neutrino fluxes usually have broad energy spectra because they are produced by pion and kaon decay in flight.
There are several exceptions using neutrinos from kaon decay at rest~\cite{PhysRevLett.120.141802,marzec2024measurementmissingenergynuclear}.
However, we cannot generate neutrinos with monochromatic energy at arbitrary energies.
\par
Electron scattering can overcome these difficulties due to the availability of many high-precision datasets with different monochromatic electron energies and scattering angles~\cite{Ankowski_2023}.
Since both electrons and neutrinos interact via electroweak interactions, their scattering processes and formalisms are similar:
Electrons interact with nucleons through vector currents, while neutrinos interact via both vector and axial vector currents.
Therefore, the same framework can be employed to simulate neutrino-nucleon and electron-nucleon scatterings.
This approach enables the validation of nuclear models used in neutrino-nucleus interaction through abundant electron scattering data.
\par
Extensions of neutrino event generators to electron scattering have been actively discussed in recent years.
\textsc{GENIE}~\cite{PhysRevD.103.113003,Khachatryan2021} has incorporated electron scattering since version 2 covering all major interaction channels: quasielastic (QE), resonance pion production, deep inelastic scattering, and meson-exchange current.
\textsc{NuWro} offers electron scattering referred to as \textsc{eWro}~\cite{eWro}.
This primarily focuses on QE while also supporting single pion production ($1\pi$) with different treatments of the nonresonance background compared to \textsc{NuWro}.
\textsc{NuWro} recently introduced an updated simulation for QE electron scattering presented in Ref.~\cite{PhysRevD.109.073004}.
A new theory-driven lepton event generator \textsc{ACHILLES}~\cite{PhysRevD.107.033007} offers neutrino and electron scatterings.
The interaction channel is currently limited to QE only.
\textsc{GiBUU}~\cite{BUSS20121} provides a unified theory and transport framework capable of simulating electron-nucleus scattering.
There are also several discussions of numerical calculation in Refs.~\cite{Bodek2019,PhysRevD.91.033005,PhysRevC.100.045503}, which are also useful for validating implementations of neutrino and electron scatterings in event generators.
\par
An implementation of electron scattering in \textsc{NEUT}~\cite{Hayato2021} was previously developed by McElwee {\it et al.}~\cite{psf2023008005,wreo32248}.
This implementation was limited to QE based on spectral function (SF) developed by Benhar {\it et al.}~\cite{BENHAR1994493,PhysRevD.72.053005}.
They evaluated the momentum-dependent removal energy correction similarly to the discussion in Sec.~\ref{sec:q3_eb}, and used it to additionally assess the uncertainty in energy transfer in the T2K analysis~\cite{Abe2023}.
However, their code was not incorporated in the distributed version of \textsc{NEUT}.
The author newly implemented electron scattering following the \textsc{NEUT} framework so that it can be integrated into the \textsc{NEUT} standard distribution.
As well as QE interactions, $1\pi$ interactions based on the DCC (dynamical coupled-channels) model ~\cite{PhysRevD.92.074024,dcc_web} is also implemented in the electron scattering framework.
This paper presents the new implementation of electron scattering in \textsc{NEUT} and the initial validations and investigations using experimental data.
\par
This paper is organized as follows;
Sec.~\ref{sec:formalism} describes the formalism of electron-nucleus scattering;
Sec.~\ref{sec:inclusive} presents comparisons with inclusive electron scattering data and extraction of the momentum-dependent removal energy correction; 
and Sec.~\ref{sec:correction} demonstrates the impact of the correction on both electron and neutrino interactions.
Note that the code developed in this study is included in the latest version of \textsc{NEUT} 5.9.0.

%%%%%%%%%%%%%%%%%%%%%%%%%%%%%
\section{Formalism} \label{sec:formalism}
The formalism of neutrino and electron scattering in \textsc{NEUT} is based on the plane wave impulse approximation (PWIA), which neglects the distortion of the struck nucleon wave function by nuclear potential.

%%%%%%%%%%%
\subsection{Quasielastic interactions based on the spectral function} \label{sec:formalism_qe}
Several nuclear models are implemented for QE interactions in \textsc{NEUT}.
A model based on the SF developed by Benhar {\it et al.}~\cite{BENHAR1994493,PhysRevD.72.053005} is extended to electron scattering.
The implementation for neutrino interactions is based on \textsc{NuWro}~\cite{JUSZCZAK2006211} with several improvements by Furmanski~\cite{warwick74036}.
The author then made several further modifications such as considering pseudoscalar form factor contribution.
Due to the similarities between neutrino and electron interactions, we can describe both interactions within the same formalism by changing the coupling constant and form factors.
A comprehensive formalism of neutrino and electron QE interactions is summarized in this section.
\par
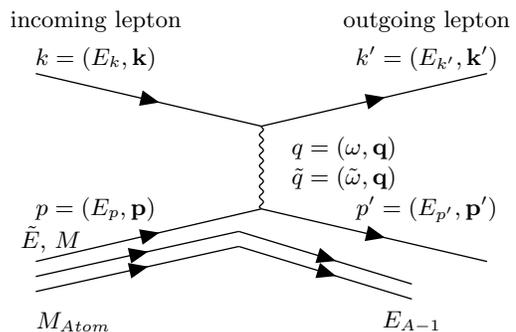
\begin{figure}[htbp] \centering
\begin{tikzpicture}
\begin{feynhand}
    \vertex (oetext1) at (2.2,3.2) {outgoing lepton};
    \vertex (ietext1) at (-2.2,3.2) {incoming lepton};
    \vertex (oetext) at (2.2,2.7) {$k'=(E_{k'},\bold{k'})$};
    \vertex (ietext) at (-2.2,2.7) {$k=(E_{k},\bold{k})$};
    \vertex (itext) at (-2.8,0.3) {$\tilde{E},\,M$};
    \vertex (itext) at (-2.2,0.7) {$p=(E_p,\bold{p})$};
    \vertex (ftext) at (2.2,0.7) {$p'=(E_{p'},\bold{p'})$};
    \vertex (inuc) at (-2.5,0.-0.8) {$M_{Atom}$};
    \vertex (fnuc) at (2,0.-0.8) {$E_{A-1}$};
    \vertex (wtext) at (1.1,1.5) {$q=(\omega,\bold{q})$};
    \vertex (wtext) at (1.1,1.1) {$\tilde{q}=(\tilde{\omega},\bold{q})$};
    \vertex [particle] (e2) at (3.0,2.5) ; % outgoing lep
    \vertex [particle] (e1) at (-3.0,2.5); % incoming lep
    \vertex [particle] (i) at (-3.0,0); % incoming nuc
    \vertex [particle] (f) at (3.0,0); % outgoing nuc
    \vertex [particle] (ir1) at (-3.0,-0.2); 
    \vertex [particle] (ir2) at (-3.0,-0.4);
    \vertex [particle] (fr1) at (2.0,-0.3); %outgoing spectator1
    \vertex [particle] (fr2) at (2.0,-0.5); %outgoing spectator2
    \vertex (w1) at (0,1.8); % W top
    \vertex (w2) at (0,0.7); % W bottom
    \vertex (wr1) at (-0.3,0.4); % spectator nucleon 1 kink
    \vertex (wr2) at (-0.3,0.2); % spectator nucleon 2 kink
    \propag [fermion] (ir1) to (wr1);
    \propag [fermion] (ir2) to (wr2);
    \propag [fermion] (wr1) to (fr1);
    \propag [fermion] (wr2) to (fr2);
    \propag [fermion] (i) to (w2);
    \propag [fermion] (w2) to (f);
    \propag [boson] (w1) to (w2);
    \propag [fermion] (w1) to (e2);
    \propag [anti fermion ] (w1) to (e1);
\end{feynhand}
\end{tikzpicture}
\caption{Diagram and variable notation for quasielastic scattering on bound nucleon.
Nucleons other than the target are treated as spectators in the assumption of the plane wave impulse approximation.
}
\label{fig:elastic}
\end{figure}

Figure~\ref{fig:elastic} shows a diagram and variable notation of QE scattering on bound nucleon.
The energy transfer $\omega$, the three-momentum transfer $\bold{q}$, and the squared of four-momentum transfer $Q^2$ can be written as follows:
\begin{equation}\begin{split}
\omega & = E_k - E_{k'}, \\ 
\bold{q} & = \bold{k-k'},\\
Q^2 & = - q^2 = - \omega^2 + \bold{q}^2.
\end{split} \end{equation}
Similarly, we can define the following parameters from nucleon kinematics:
\begin{equation}\begin{split}
\tilde{\omega} & = E_{p'}-E_p, \label{eq:omegatilde}\\
\tilde{Q}^2 & = -\tilde{q}^2 = -\tilde{\omega}^2+(\bold{p'-p})^2 
  = -\tilde{\omega}^2 + \bold{q}^2, \\
E_p &= \sqrt{M^2 + \bold{p}^2}.
\end{split} \end{equation}
Note that the energy of the target nucleon $E_p$ is calculated using the on-shell nucleon mass $M$.
This formulation is based on the de Forest approximation~\cite{DEFOREST1983232}, which assumes that the target nucleon has on-shell mass $M$, while the outgoing nucleon gains only a part of the energy transfer, as also discussed in Ref.~\cite{PhysRevC.86.024616}.
From the energy conservation, we can get
\begin{align}
 E_k + M_{Atom} & = E_{k'} + E_{A-1} + E_{p'},\label{eq:cons_ene}
\end{align}
where $M_{Atom}$ is the target nucleus mass and $E_{A-1}$ is the energy of residual nucleus.
The removal energy $\tilde{E}$ can be defined as 
\begin{align}
\tilde{E} & = E_k - E_{k'} -E_{p'}+M = \omega +M -E_{p'} \label{eq:etilde}.
\end{align}
%where $M$ is the nucleon mass.
We can get the momentum-dependent binding energy $\epsilon_B$:
\begin{equation}\begin{split}
  \epsilon_B = \omega - \tilde{\omega} = \tilde{E} + E_p - M.
\end{split} \end{equation}

The total cross section is given by the following equation under the assumption of the PWIA:
\begin{align}
\frac{d\sigma_\text{tot}}{dQ^2}
& = \int d^3p\,d\tilde{E}\,P_{\text{hole}}(\bold{p},\tilde{E}) \frac{d\sigma}{dQ^2}, \label{eq:tot}
\end{align}
where $P_{\text{hole}}(\bold{p},\tilde{E})$ represents the probability of removing a nucleon with momentum $\bold{p}$ and removal energy $\tilde{E}$, given by the SF.
And $d\sigma/dQ^2$ denotes the elementary cross section:
\begin{align}
\frac{d\sigma}{dQ^2}
& = \frac{C}{E_k}
\int d^3k'\,\delta(\omega + M  - \tilde{E}-E_{p'})
\frac{L_{\mu\nu}H^{\mu\nu}}{E_p E_{p'}E_{k'}},\label{eq:ele_furmanski}
\end{align}
where the $L_{\mu\nu}$ and $H^{\mu\nu}$ represent leptonic and hadronic tensors, respectively.
The coupling constant $C$ can be written as follows for charged-current (CC), neutral-current (NC), and electromagnetic (EM) interactions:
\begin{align}
C &=
\begin{dcases}
 \frac{G_F^2\cos^2{\theta_C}}{8\pi^2} & (\text{CC}),\\
 \frac{G_F^2}{8\pi^2} & (\text{NC}),\\
 \frac{\alpha^2}{Q^4} & (\text{EM}),
\end{dcases}
\end{align}
where $G_F$, $\alpha$, and $\theta_{C}$ denote the Fermi constant, the fine structure constant, and the Cabbibo angle, respectively.
Since the massless photon mediates the EM, the factor of $1/Q^4$ appears.
The elementary cross section includes the integral of outgoing lepton momentum $d^3k'$.
In \textsc{NEUT} event generation procedure, it is converted to the center-of-mass variables as described in Appendix~\ref{sec:qe_neut}.
\par
The leptonic and hadronic tensors $L_{\mu\nu}H^{\mu\nu}$ can be written as 
\begin{align} 
L_{\mu\nu}H^{\mu\nu} = 2 \sum_{j=1}^5 A_j W_j,
\end{align} 
where $A_j$ is given by
\begin{equation}\begin{split} \label{eq:A1_5} 
A_1 &= 2M^2 k \cdot k', \\
A_2 &= 2p\cdot k~p\cdot k' - p^2 k\cdot k',\\ 
A_3 &= k\cdot\tilde{q}~k'\cdot p - k\cdot p~k'\cdot \tilde{q},\\
A_4 &= k\cdot k'~\tilde{q}^2 - 2k\cdot \tilde{q}~k'\cdot\tilde{q},\\
A_5 &= k\cdot p ~k'\cdot\tilde{q} + k'\cdot p~k\cdot\tilde{q} 
 - k\cdot k'~ p\cdot \tilde{q},
\end{split} \end{equation}
and $W_j$ can be expressed as
\begin{equation}\begin{split} \label{eq:W_1_5}
W_1 &= (F_A)^2(1+\tau) + \tau(F_1+F_2)^2,\\ 
W_2 &= (F_A)^2 + (F_1)^2 + \tau \left[ (F_2)^2+4(F_{3A})^{2} \right],\\ 
W_3 &= \pm2F_A(F_1+F_2),\\
W_4 &= \frac{1}{4} \left[ (F_2)^2 -\tau(F_2 - 2F_{3V})^2 - 4\tau(F_P+F_{3A})^2 \right] \\ 
 &- (F_{3V})^2 - \frac{1}{2}\left[F_1(2F_{3V}-F_2) - 2F_A(F_P+F_{3A})\right],\\ 
W_5 &= W_2 \pm 2\left[F_{3V}(F_1-\tau F_2)- F_{3A}(F_A-2\tau F_P) \right],
\end{split} \end{equation}
with $\tau = -\tilde{q}^2/(4M^2)$ (Ref.~\cite{PhysRevC.86.024616}).
The upper (lower) sign corresponds to neutrino (antineutrino).
The second-class form factors are tuned off by default: $F_{3A}=F_{3V}=0$.
he axial vector form factor $F_A$ and the pseudoscalar form factor $F_P$ for CC, NC, and EM interactions can be expressed as follows, assuming dipole parametrization (Ref.~\cite{ALBERICO2002227}):
\begin{equation}\begin{split} \label{eq:axial_pseudo}
F_A^{\text{CC}} & = g_A \left( 1+ \frac{\tilde{Q}^2}{M_A^2} \right)^{-2}, \\
F_A^{\text{NC}} & = \frac{1}{2}(\pm F_A^\text{CC} + F_A^s) \\
 & = \frac{1}{2} (\pm g_A + g_A^s) \left( 1+ \frac{\tilde{Q}^2}{M_A^2} \right)^{-2}, \\
F_P^{\text{CC,NC}} & = \frac{2 M^2}{m_\pi^2 + \tilde{Q}^2} F_A^{\text{CC,NC}},\\
F_A^{\text{EM}} &= F_P^{\text{EM}} = 0, 
\end{split} \end{equation}
where the axial coupling constant is $g_A=-1.2673$ and the axial mass is $M_A=1.21$\,GeV by default. 
The upper (lower) sign corresponds to the proton (neutron) form factors.
The strange axial coupling constant $g_A^s$ is known to prefer negative values~\cite{PhysRevC.78.015207,PhysRevD.84.014002,PhysRevD.88.093004,PhysRevD.107.072006}.
\textsc{NEUT} sets $g_A^s = 0$ by default, neglecting the strange quark contribution. 
The vector form factors for CC, NC, and EM interactions can be written as 
\begin{equation}\begin{split} \label{eq:VFF}
F_{1,2}^{\text{CC}} & = F_{1,2}^p - F_{1,2}^n,\\
F_{1,2}^{\text{NC},p/n} & =\pm \frac{1}{2}F_{1,2}^{\text{CC}} - 2\sin^2\theta_W F_{1,2}^{p/n}, \\ 
F_1^{p/n} & = \frac{G_E^{p/n} + \tau G_M^{p/n}}{1+\tau},\\
F_2^{p/n} & = \frac{G_M^{p/n} - G_E^{p/n}}{1+\tau},\\ 
F_1^{\text{EM},p/n} & = \frac{G_E^{p/n} + \tau G_M^{p/n}}{1+\tau},\\
F_2^{\text{EM},p/n} & = \frac{G_M^{p/n} - G_E^{p/n}}{1+\tau},\\
\end{split} \end{equation}
where $G_E^{p/n}$ and $G_M^{p/n}$ are Sachs form factors, and $\theta_W$ is the Weinberg angle.
The upper (lower) sign of the NC form factor corresponds to the proton (neutron).
The strangeness contributions of the vector form factors are consistent with zero~\cite{PhysRevC.78.015207}, thus it is neglected in Eq.~(\ref{eq:VFF}).
%%%%
There are several parametrizations for the Sachs form factors.
A simple dipole parametrization can be written as
\begin{equation}\begin{split} \label{eq:VFF_di}
\frac{G_M^p}{\mu_p} & = \frac{G_M^n}{\mu_n} = G_E^p = D,\\
D & = \left( 1+\frac{\tilde{Q}^2}{M_V^2} \right)^{-2}, \\
G_E^n & = 0,
\end{split} \end{equation}
where $\mu_p=2.793$ and $\mu_n=-1.913$ for the magnetic moments of proton and neutron~\cite{ParticleDataGroup:2024cfk}, and $M_V=0.84$\,GeV for vector mass.
For subsequent calculations using \textsc{NEUT}, a more sophisticated parametrization, BBBA05~\cite{BRADFORD2006127}, derived from electron scattering data, is employed.
The tensors and form factors in Eqs.~(\ref{eq:A1_5})- (\ref{eq:VFF_di}) use $\tilde{Q}^2$ and $\tilde{q}$ instead of $Q^2$ and $q$.
This substitution arises as a result of the de Forest approximation mentioned earlier.
\par
Pauli blocking is implemented using a simple step function, which requires the momentum of the outgoing nucleon to exceed the Fermi surface $p_F$: $|\bold{p'}|>p_{F}$.
\textsc{NEUT} uses $p_{F}=209$\,MeV for $^{12}$C and $^{16}$O by default, which is the average Fermi momentum calculated from nuclear density distributions~\cite{PhysRevD.72.053005}, 
In the case of CC and EM interactions, the Coulomb potential $|V_{eff}|$ takes effect.
\textsc{NEUT} currently considers this effect only for EM.
The effect of the Coulomb potential is considered by 
\begin{equation}\begin{split}
E_k & = E_0 + |V_{eff}|, \\
E'_k  &= E'_0+ |V_{eff}|,
\end{split} \end{equation}
where $E_0$ $(E'_0)$ is the incoming (outgoing) electron energy at a sufficient distance from the nucleus.
\textsc{NEUT} sets a constant value of $|V_{eff}|= 3.1$\,MeV and 3.4\,MeV for $^{12}$C and $^{16}$O, respectively~\cite{PhysRevC.60.044308,Bodek2019}.
\par
\begin{figure}[htbp] \centering
\begin{minipage}[b]{0.85\columnwidth} \centering
\includegraphics[width=1.0\textwidth]{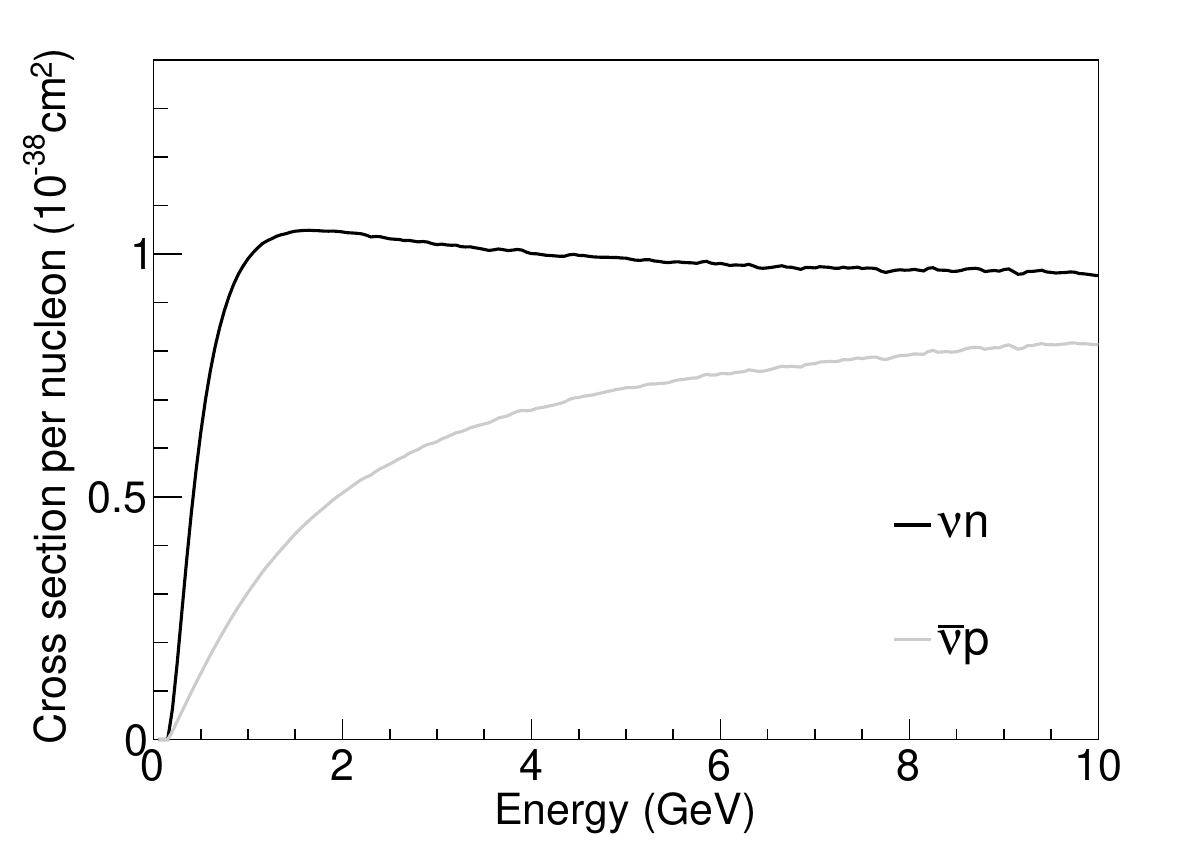}
\end{minipage}
\\
\begin{minipage}[b]{0.85\columnwidth} \centering
\includegraphics[width=1.0\textwidth]{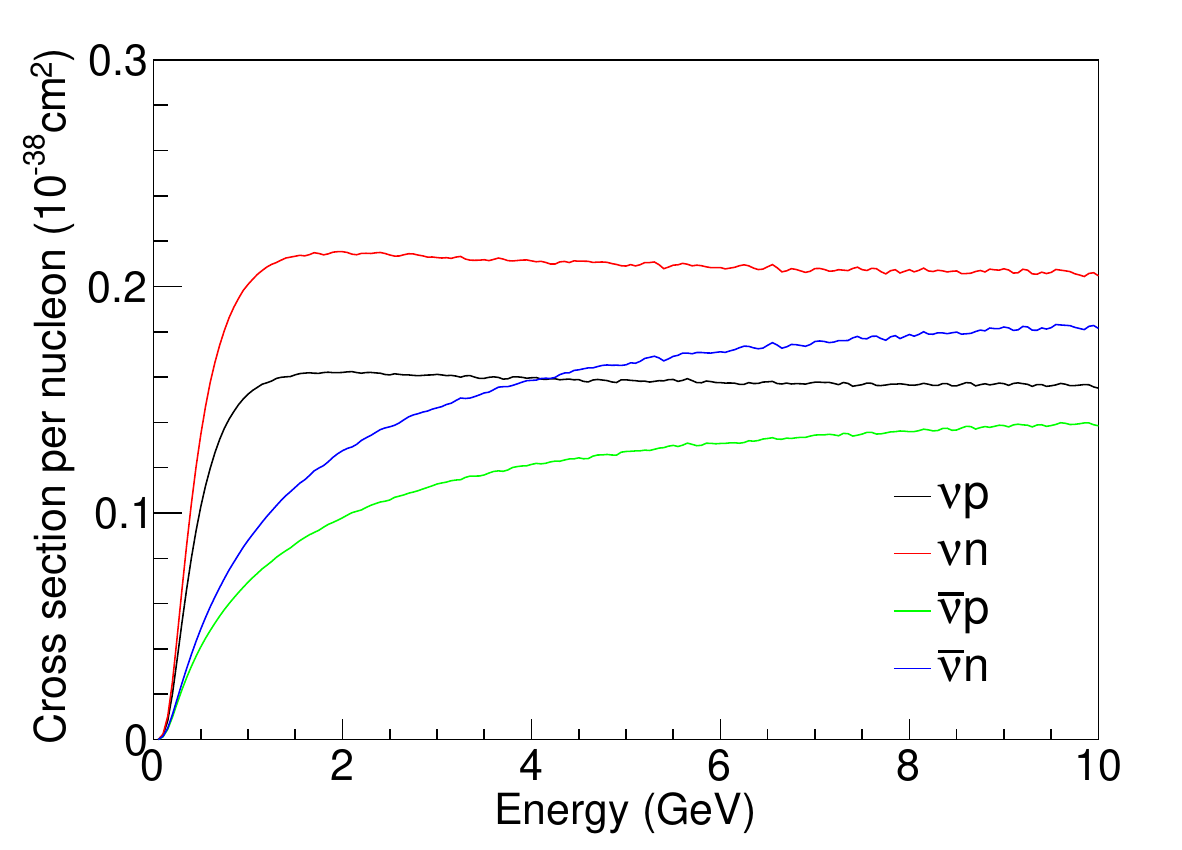}
\end{minipage}
\\
\begin{minipage}[b]{0.85\columnwidth} \centering
\includegraphics[width=1.0\textwidth]{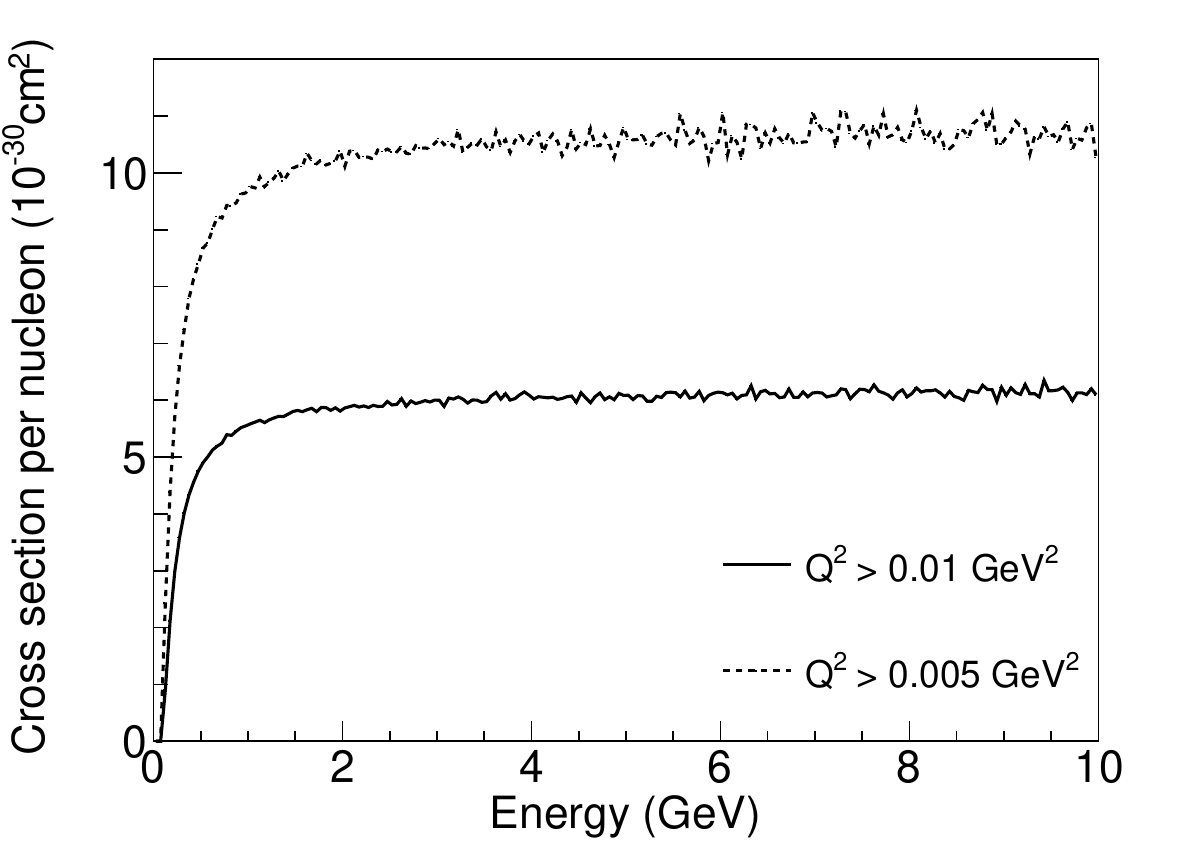}
\end{minipage}
\caption{Quasielastic cross section per nucleon as a function of incoming lepton energy of CC (top), NC (middle), and EM (bottom) interactions.
The target nucleus is $^{16}$O for CC and NC interactions and $^{12}$C for EM interactions.
Fermi momentum of $p_F=209$\,MeV is employed in calculations.
The EM cross section is calculated with two different $Q^2$ cutoffs: 0.01\,GeV$^2$ and 0.005\,GeV$^2$.
\textsc{NEUT} provides the table calculated with $Q^2 > 0.01$\,GeV$^2$.
}
\label{fig:xsec}
\end{figure}

Figure~\ref{fig:xsec} shows QE cross sections of CC, NC, and EM interactions.
\textsc{NEUT} employs a cutoff of $Q^2>0.01$\,GeV$^2$ in Eq.~(\ref{eq:tot}) for EM interactions to increase simulation speed, which is sufficiently small for this study.
The EM cross section is largely affected  by this cutoff threshold.
As a reference, if a cutoff of $Q^2>0.005$\,GeV$^2$ is adopted, the cross section increases approximately 1.6 times.
It is worth noting that the same $Q^2$ cutoff must be employed in the event generation, since \textsc{NEUT} reads a precomputed cross section, i.e., it is not an on-the-fly calculation.
The NCQE model explained here is different from that discussed in the past Super-Kamiokande analysis~\cite{PhysRevD.99.032005,PhysRevD.109.L011101,Harada_2023}.
They used the cross section calculated with different parameters $g_A^s=-0.08$~\cite{ALEXAKHIN20078} and different code written by the authors of Refs.~\cite{Ueno:2012tpn,PhysRevLett.108.052505}.
When $g_A^s=-0.08$ is used in this model,  good consistency in the cross section is observed between this model and their model.
The cross sections have statistical fluctuation arising from the finite number of calculation trials explained in Appendix~\ref{sec:qe_neut}.
In event generation, these cross section data are interpolated to mitigate its impact on the simulation.

When the target nucleon is unbound, $\tilde{q}$ and $q$ become equivalent.
Then, $A_4$ can be written as
\begin{align}
A_4 = \frac{1}{2} (q^2-m^2)m^2.
\end{align}
We obtain $A_4=0$ in the case of NC ($m\rightarrow0$).
The pseudoscalar form factor $F_P$ only appears in $W_4$ when the second-class form factors are absent.
Therefore, in the case of an unbound nucleon target, the pseudoscalar form factor $F_P$ does not contribute to the NC cross section.
The same conclusion can be confirmed by the Llewellyn--Smith formalism~\cite{LLEWELLYNSMITH1972261}, which describes interaction with unbound nucleon.
In contrast, for a bound nucleon, $A_4$ has a nonzero value because of the presence of removal energy $\tilde{E}$.
Consequently, the pseudoscalar form factor $F_P$ contributes to the NC cross section, approximately $3\%$.
\par
The SF $P_{\text{hole}}(\bold{p},\tilde{E})$ consists of two components: a mean-field term and a correlated term.
In \textsc{NEUT} framework, the mean-field term corresponds to $|\bold{p}|<300$\,MeV and $\tilde{E}<100$\,MeV, while the correlated term corresponds to the remaining phase space.
It should be noted that the correlated term also contributes to $|\bold{p}|<300$\,MeV and $\tilde{E}<100$\,MeV as shown in Ref.~\cite{PhysRevC.110.054612}.
However, \textsc{NEUT} currently does not distinguish this contribution from the mean-field term.
The correlated term gives a tail in momentum and removal energy, contributing approximately 20\% to the total.
Interactions involving a correlated nucleon are considered to be accompanied by the emission of an additional paired nucleon.
The nucleon pair ratios ($pp,\,nn,\,pn$) have been measured in electron scattering experiments~\cite{doi:10.1126/science.1156675,PhysRevLett.97.162504}.
Approximately $90\pm10$\% of correlated pairs are $pn$ while $pp$ and $nn$ have equal contributions of $5\pm1.5$\% under the assumption of isospin symmetry.
\textsc{NEUT} assumes 100\% of the correlated pairs are $pn$ by default.
The struck nucleon, along with any correlated nucleon pairs, subsequently undergoes the final-state interactions (FSI), which is described by the intranuclear cascade model explained in Ref.~\cite{Hayato2021}.
\textsc{NEUT} provides two nuclear deexcitation models.
One is a simple data-driven model based on Ref.~\cite{kobayashi2006deexcitation} employed by default.
The other is a dedicated nuclear deexcitation event generator \textsc{NucDeEx}~\cite{PhysRevD.109.036009,code,Abe_2021}.
Radiative corrections are not implemented for electron scattering, while it is approximately implemented for charged-current quasielastic (CCQE) interaction following Refs.~\cite{Tomalak2022,PhysRevD.106.093006}.
In the future, radiative correction could be incorporated, for example, through the universal implementation of radiative correction introduced in Ref.~\cite{TENAVIDAL2025109509}.

%%%%%%%%%%%
\subsection{Single pion production based on the DCC model} \label{sec:formalism_1pi}
An electroweak meson production reaction model, the DCC model~\cite{PhysRevC.100.045503, PhysRevD.92.074024}, has been implemented for electron scattering in \textsc{NEUT} as a part of this study.
The code provided by the DCC authors was initially implemented for neutrino interactions by Yamauchi et al.~\cite{Yamauchi:2024lby} and was subsequently extended to electron scattering by the author.
Although the DCC model can describe double pion production, only $1\pi$ interaction is considered in these implementations.
\par
\begin{figure}[htbp] \centering
\includegraphics[width=0.90\columnwidth]{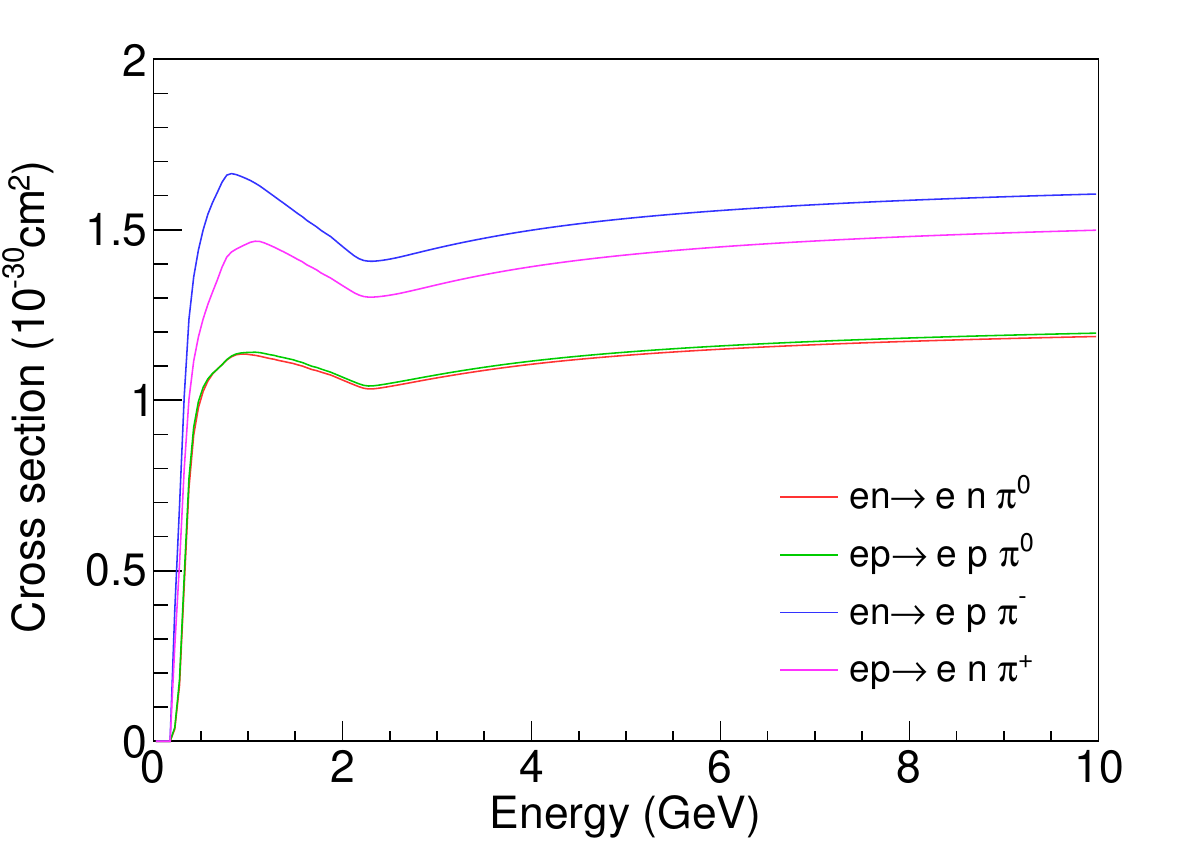}
\caption{Single pion production cross section for electrons based on the DCC model as a function of electron energy~\cite{PhysRevC.100.045503,PhysRevD.92.074024}.
The cross section is calculated with $W<2.1$\,GeV.}
\label{fig:xsec_1pi}
\end{figure}

For both neutrino and electron interactions, \textsc{NEUT} assumes that the target nucleon moves with momentum below the Fermi surface with on-shell mass.
The nucleon momentum is assumed to be uniformly distributed in the momentum phase space.
Interactions are calculated in the nucleon rest frame, and the particle kinematics are then boosted to the laboratory frame.
\textsc{NEUT} currently does not account for removal energy.
For example, the T2K analysis addresses this limitation by using \textsc{NuWro} to estimate the effect of removal energy, which is then incorporated into the oscillation analysis.
Pauli blocking can be optionally considered using a step-function, similar to that used in QE.
In this study, the invariant mass is restricted to $W<2.1$\,GeV.
Figure.~\ref{fig:xsec_1pi} shows the $1\pi$ cross section for electrons according to the DCC model.

%%%%%%%%%%%%%%%%%%%%%%%%%%%%%
\section{Comparison with inclusive electron scattering data} \label{sec:inclusive}
This section discusses comparisons between \textsc{NEUT} predictions and inclusive electron scattering data on carbon and oxygen, $^{12}$C$(e,e')$ and $^{16}{\rm O}(e,e')$.
The \textsc{NEUT} implementation of the QE model was first validated by comparing it with numerical calculations by Ankowski {\it et al.}~\cite{PhysRevD.91.033005}.
The \textsc{NEUT} simulations were found to be consistent with their numerical calculations, confirming the validity of the implementation.
The impact of Coulomb potential and Pauli blocking on \textsc{NEUT} simulations was also investigated.
Both effects were found to be small.
Details of these foundational studies are provided in Appendix~\ref{sec:appen_qe}.
\par
\begin{figure*}[htbp] \centering
\begin{minipage}[b]{0.495\textwidth} \centering
\includegraphics[width=1.0\textwidth]{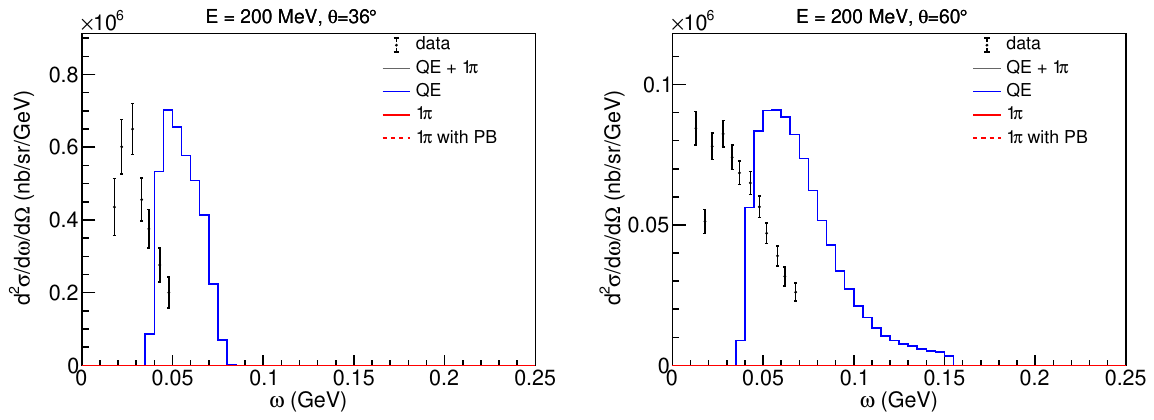}
\end{minipage}
\begin{minipage}[b]{0.495\textwidth} \centering
\includegraphics[width=1.0\textwidth]{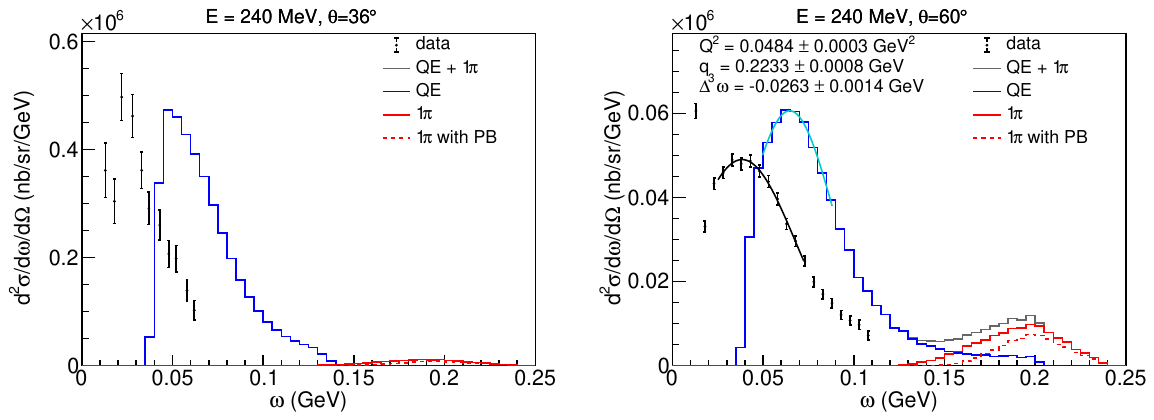}
\end{minipage}
\\
\begin{minipage}[b]{0.495\textwidth} \centering
\includegraphics[width=1.0\textwidth]{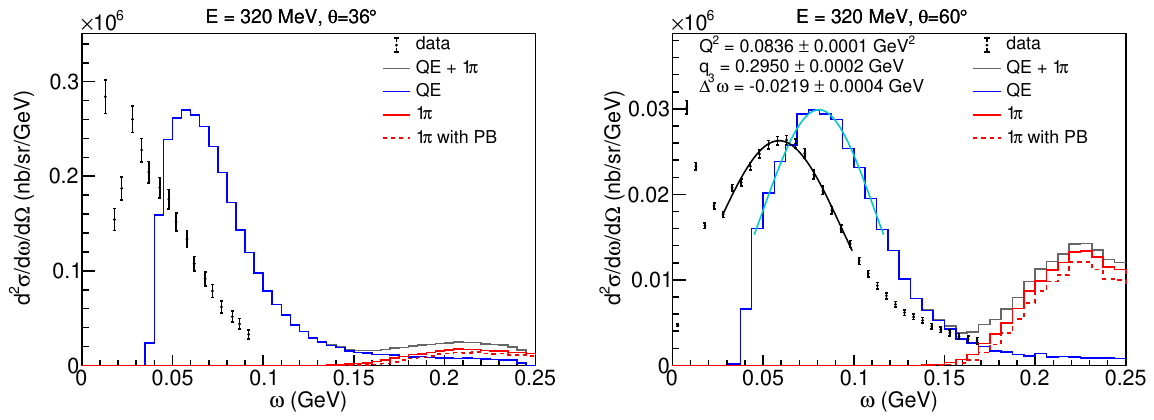}
\end{minipage}
\begin{minipage}[b]{0.495\textwidth} \centering
\includegraphics[width=1.0\textwidth]{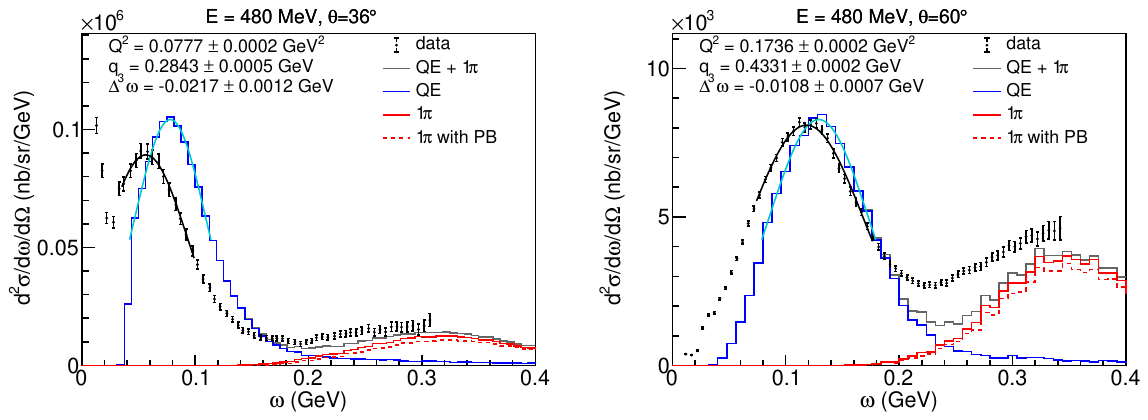}
\end{minipage}
\\
\begin{minipage}[b]{0.495\textwidth} \centering
\includegraphics[width=1.0\textwidth]{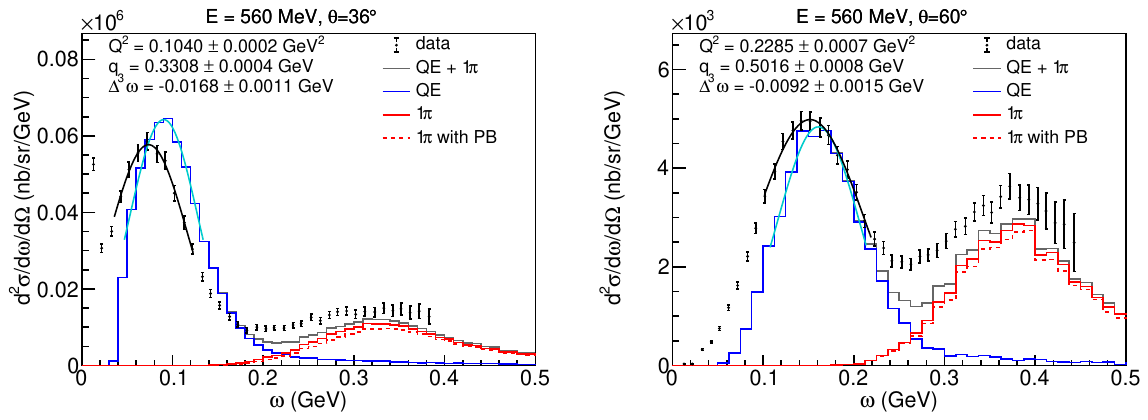}
\end{minipage}
\begin{minipage}[b]{0.495\textwidth} \centering
\includegraphics[width=1.0\textwidth]{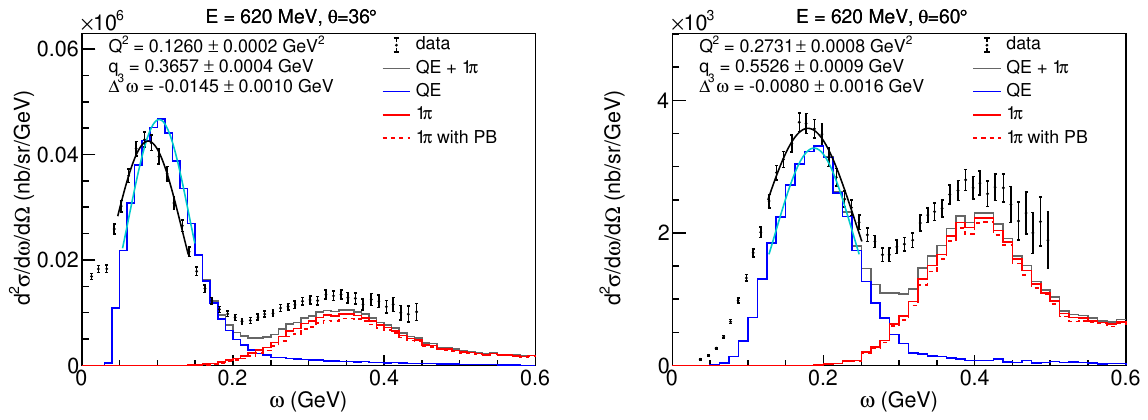}
\end{minipage}
\\
\begin{minipage}[b]{0.495\textwidth} \centering
\includegraphics[width=1.0\textwidth]{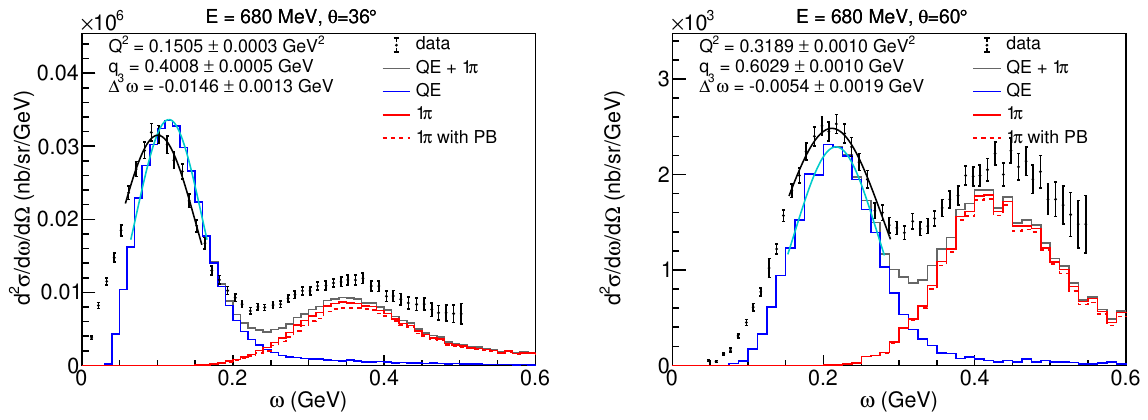}
\end{minipage}
\hspace{0.495\textwidth}
\caption{
Comparisons of inclusive electron scattering $^{12}$C$(e,e')$ cross section between \textsc{NEUT} and experimental data~\cite{Barreau:1983ht}.
The figures are labeled by electron energy and scattering angle.
The blue line shows the contribution of quasielastic and the red solid (dashed) line shows the contribution of single pion production without (with) Pauli blocking effect.
The gray line shows the total cross section, summing QE and single pion production without Pauli blocking effect.
The black (cyan) curved line on the QE peak shows the result of the Gaussian fitting discussed in Sec.~\ref{sec:q3_eb}.
Parameters of $Q^2,\,q_3,\,$ and $\Delta \omega$ written in the figures are defined as Eqs.~(\ref{eq:domega}) and (\ref{eq:Q2_q3}).
}
\label{fig:Barreau}
\end{figure*}
\begin{figure*}[htbp] \centering
\begin{minipage}[t]{0.495\textwidth} \centering
\includegraphics[width=1.0\textwidth]{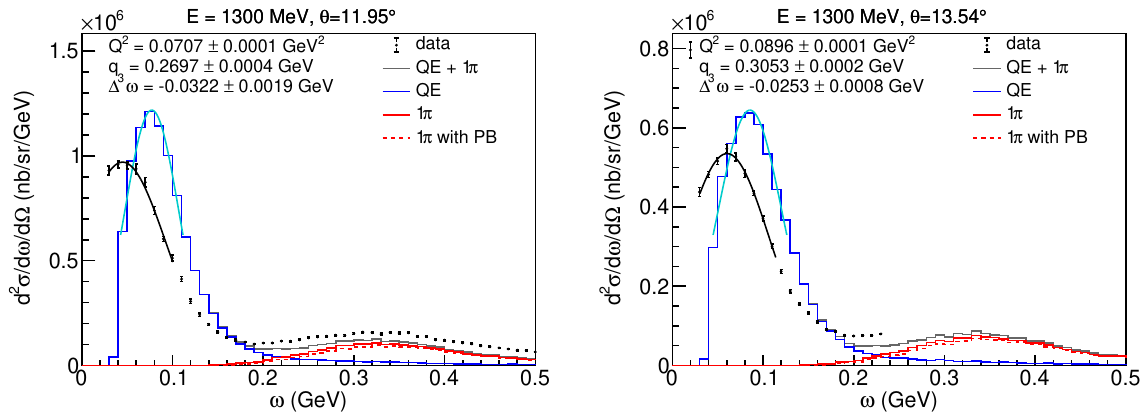}
\end{minipage}
\begin{minipage}[t]{0.495\textwidth} \centering
\includegraphics[width=1.0\textwidth]{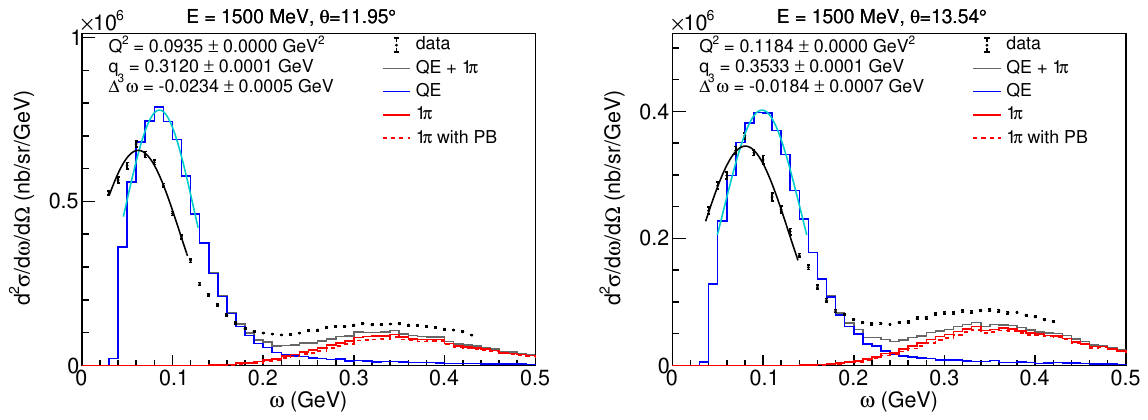}
\end{minipage}
\\
\begin{minipage}[b]{0.495\textwidth} \centering
\includegraphics[width=1.0\textwidth]{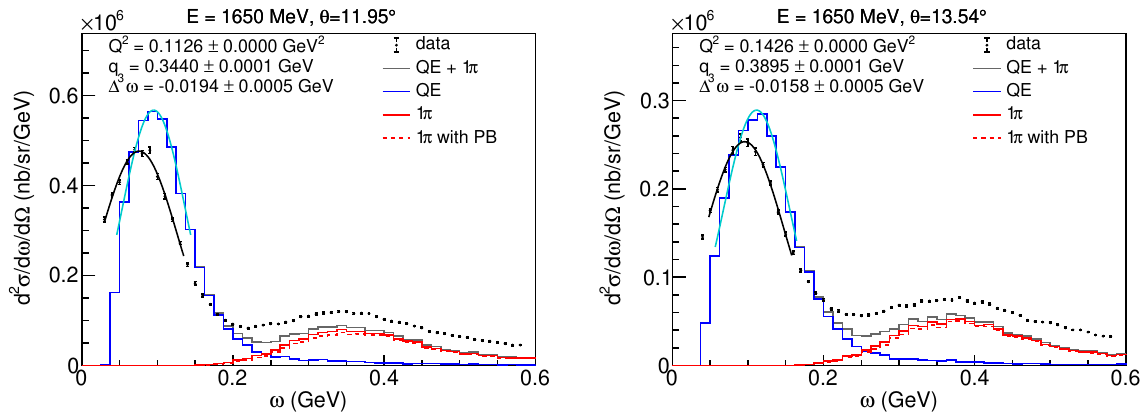}
\end{minipage}
\hspace{0.495\textwidth}
\caption{
  The same as Fig.~\ref{fig:Barreau} but comparisons with different experimental data~\cite{Baran:1988tw}.
}
\label{fig:Baran}
\end{figure*}
\begin{figure*}[htb] \centering
\begin{minipage}[t]{0.49\columnwidth} \centering
\includegraphics[width=1.0\textwidth]{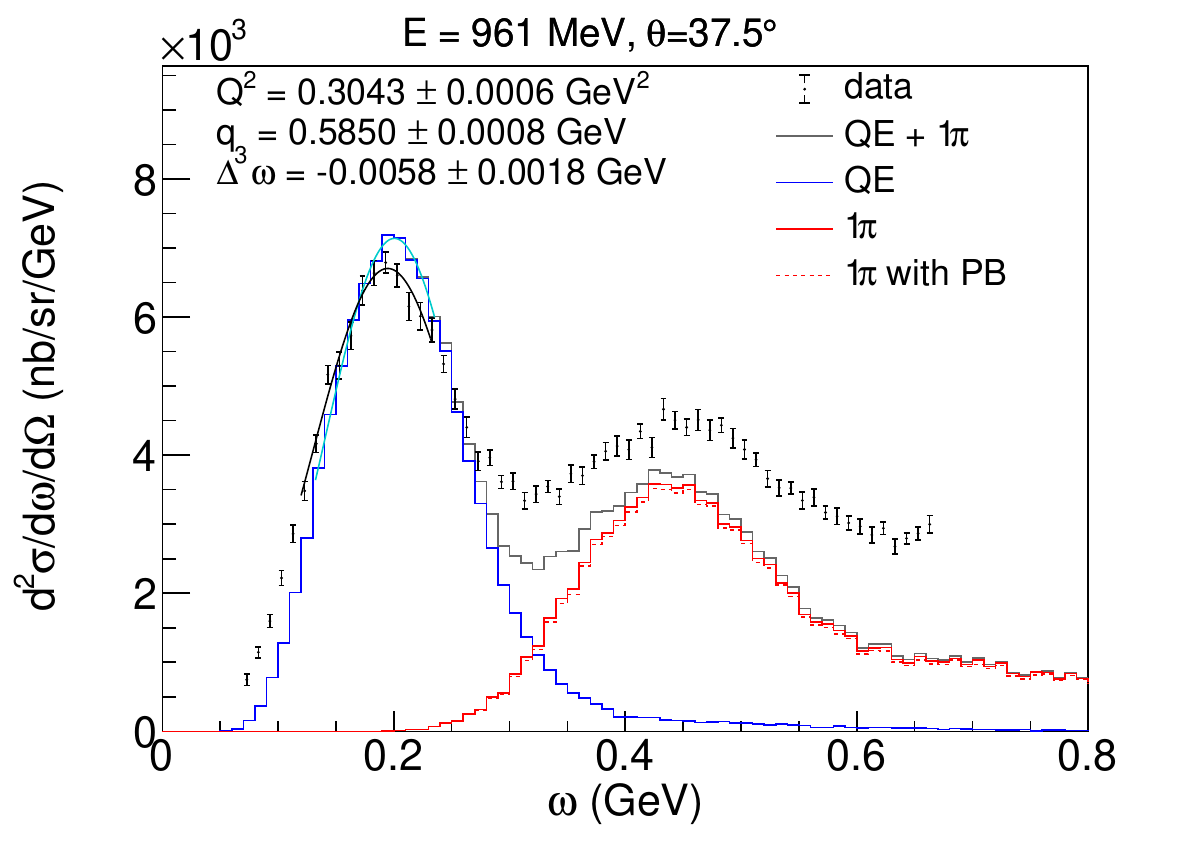}
\end{minipage}
\begin{minipage}[t]{0.49\columnwidth} \centering
\includegraphics[width=1.0\textwidth]{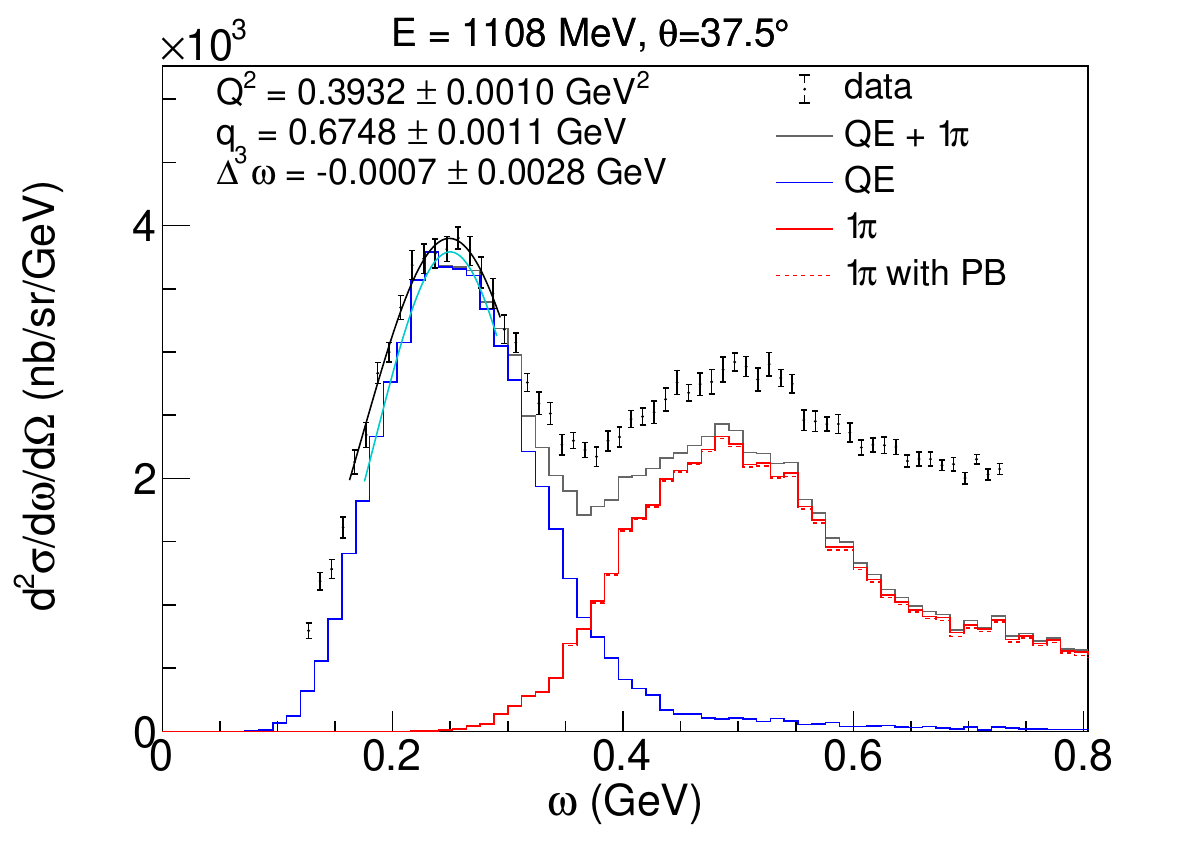}
\end{minipage}
\begin{minipage}[b]{0.49\columnwidth} \centering
\includegraphics[width=1.0\textwidth]{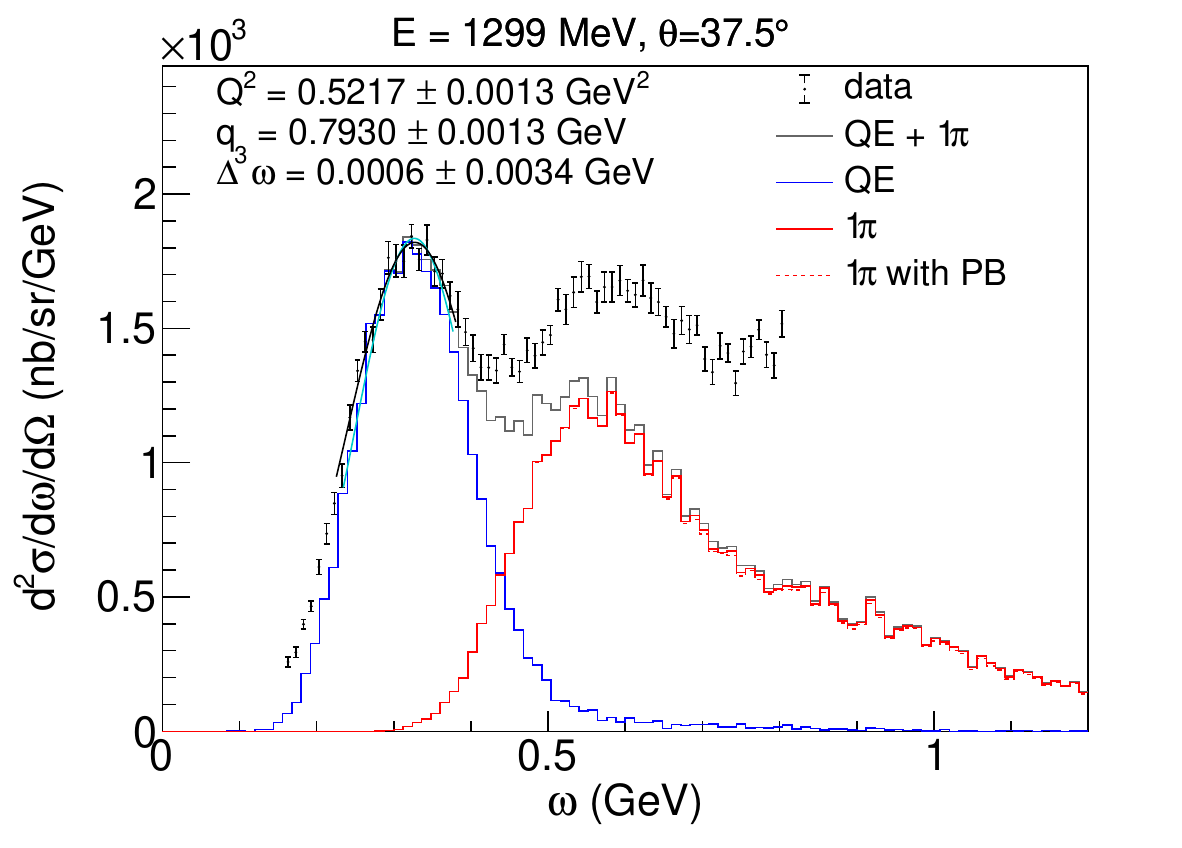}
\end{minipage}
\begin{minipage}[b]{0.49\columnwidth} \centering
\includegraphics[width=1.0\textwidth]{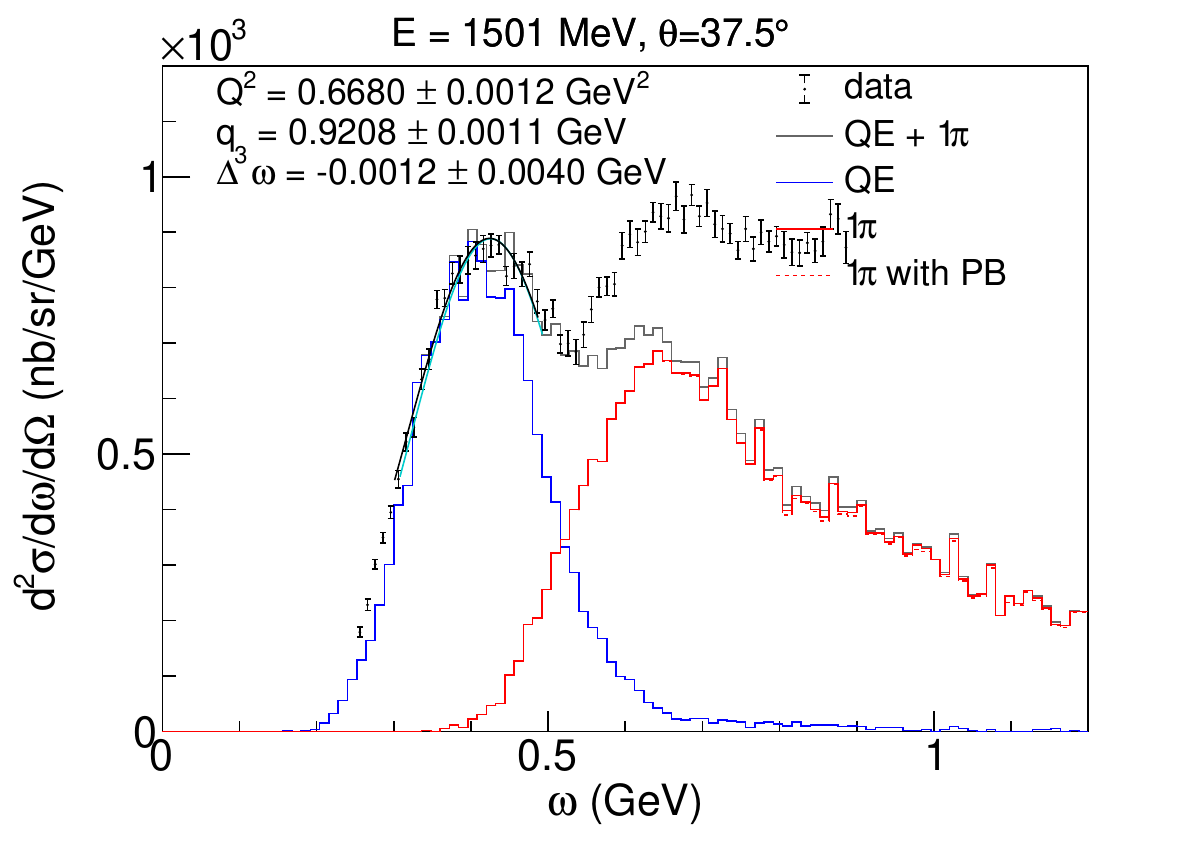}
\end{minipage}
\caption{
  The same as Fig.~\ref{fig:Barreau} but comparisons with different experimental data~\cite{Sealock:1989nx}.
}
\label{fig:Sealock}
\end{figure*}
\begin{figure*}[htb] \centering
\begin{minipage}[t]{1.0\columnwidth} \centering
\includegraphics[width=1.0\textwidth]{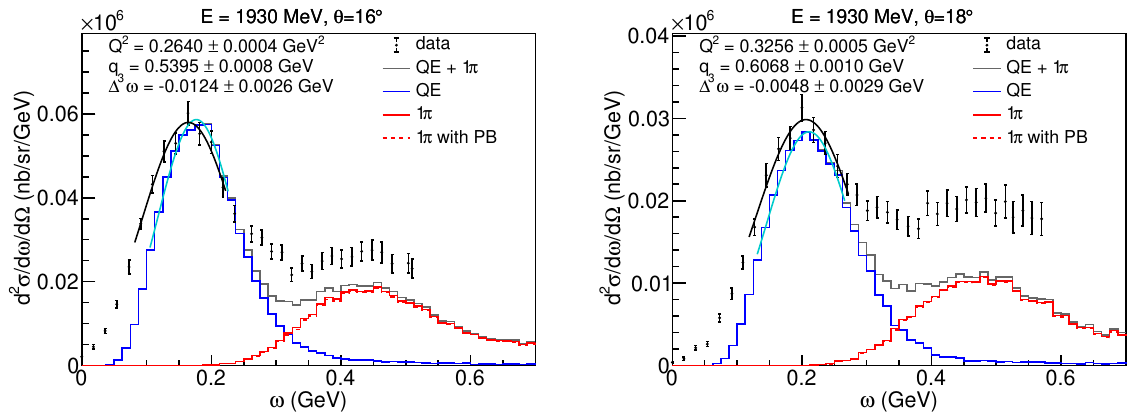}
\end{minipage}
\begin{minipage}[t]{1.0\columnwidth} \centering
\includegraphics[width=1.0\textwidth]{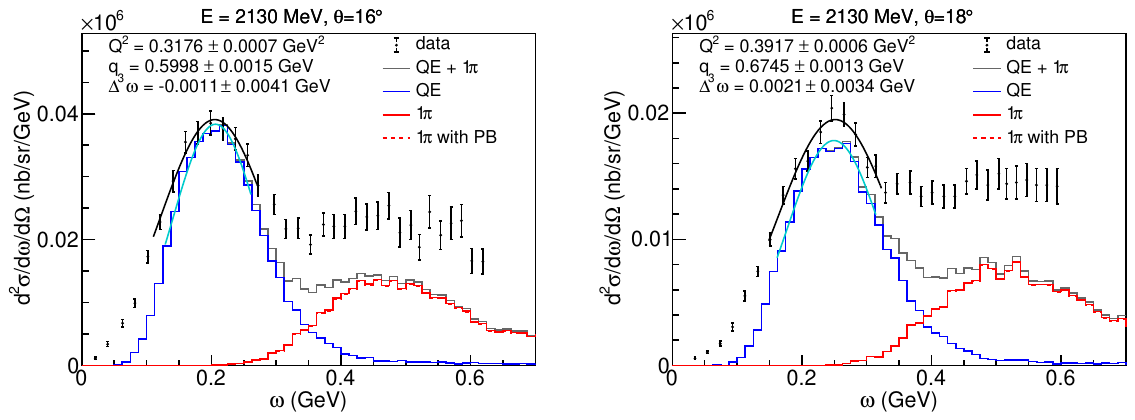}
\end{minipage}
\caption{
  The same as Fig.~\ref{fig:Barreau} but comparisons with different experimental data~\cite{Bagdasaryan:1988hp}.
}
\label{fig:Bagdasaryan}
\end{figure*}
\begin{figure*}[htb] \centering
\includegraphics[width=1.0\columnwidth]{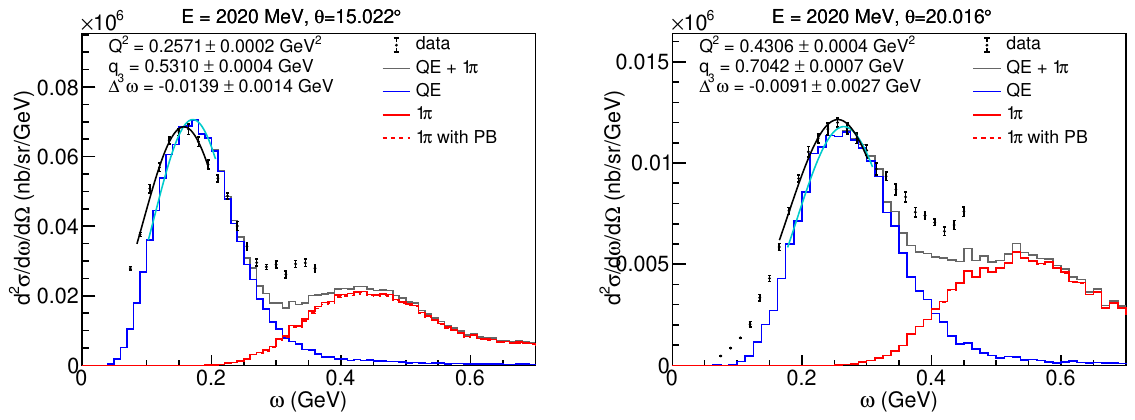}
\caption{
  The same as Fig.~\ref{fig:Barreau} but comparisons with different experimental data~\cite{Day:1993md}.
}
\label{fig:Day}
\end{figure*}
\begin{figure*}[htbp] \centering
\begin{minipage}[t]{0.30\textwidth} \centering
\includegraphics[width=1.0\textwidth]{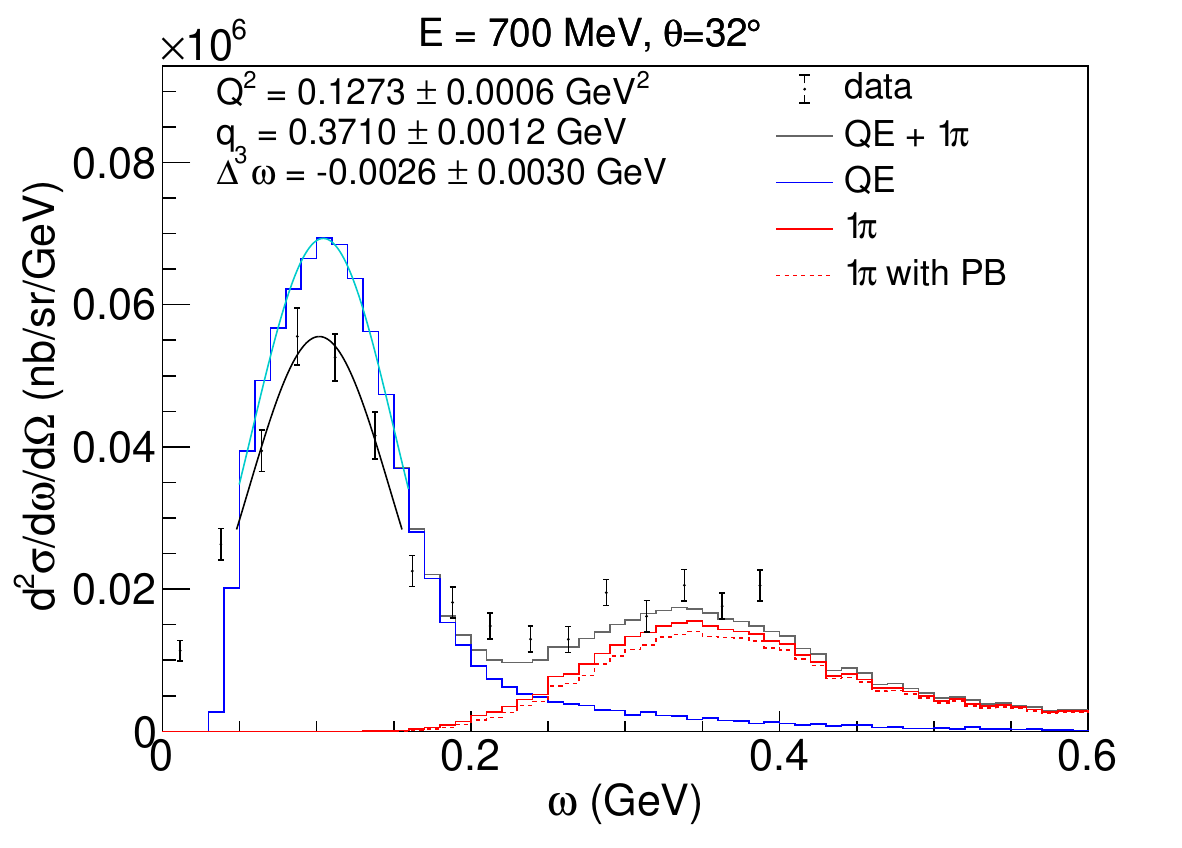}
\end{minipage}
\begin{minipage}[t]{0.30\textwidth} \centering
\includegraphics[width=1.0\textwidth]{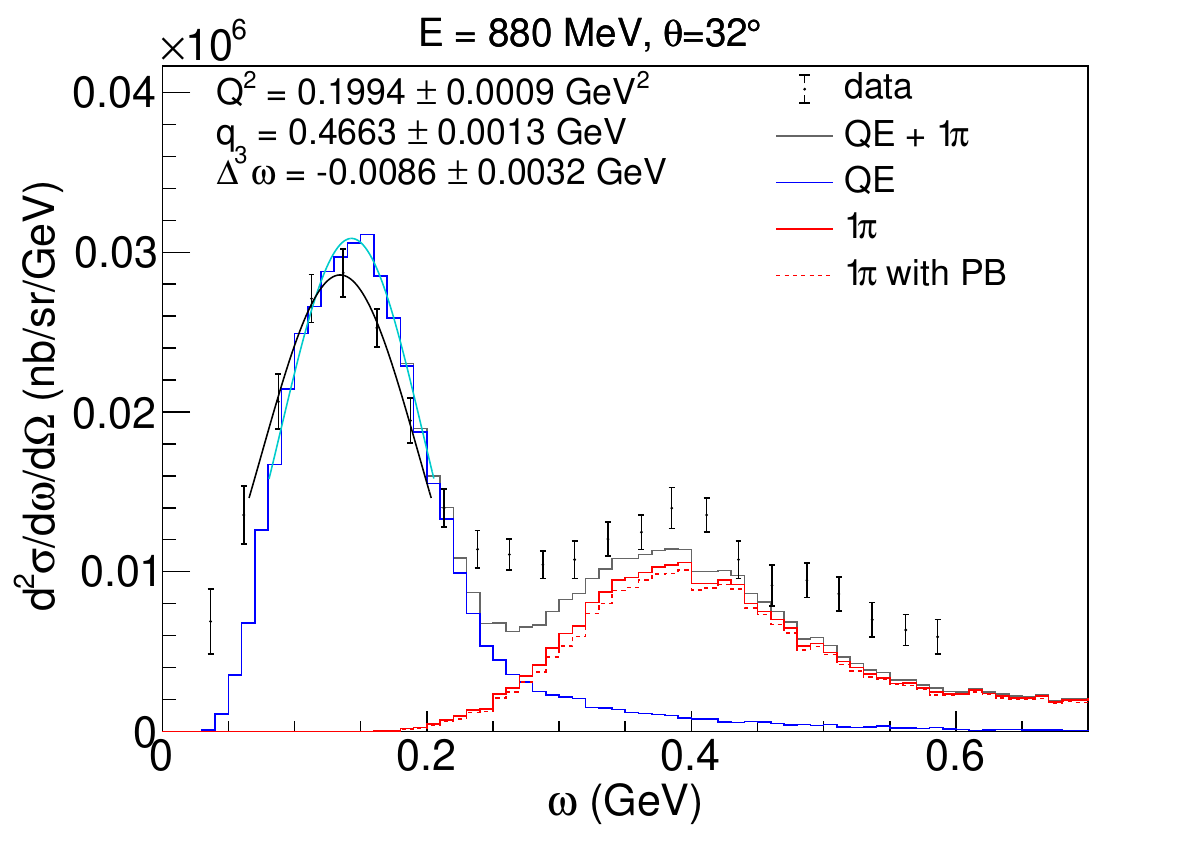}
\end{minipage}
\begin{minipage}[t]{0.30\textwidth} \centering
\includegraphics[width=1.0\textwidth]{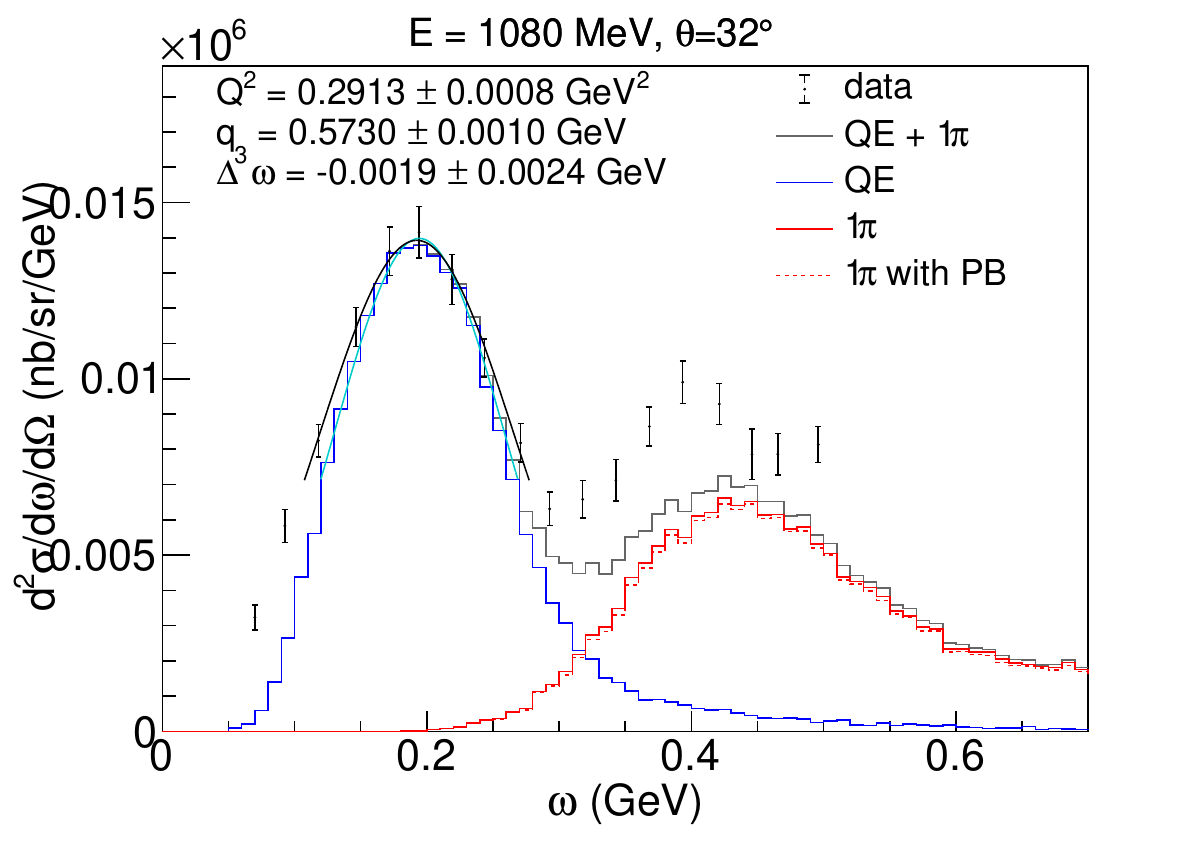}
\end{minipage}
\\
\begin{minipage}[t]{0.30\textwidth} \centering
\includegraphics[width=1.0\textwidth]{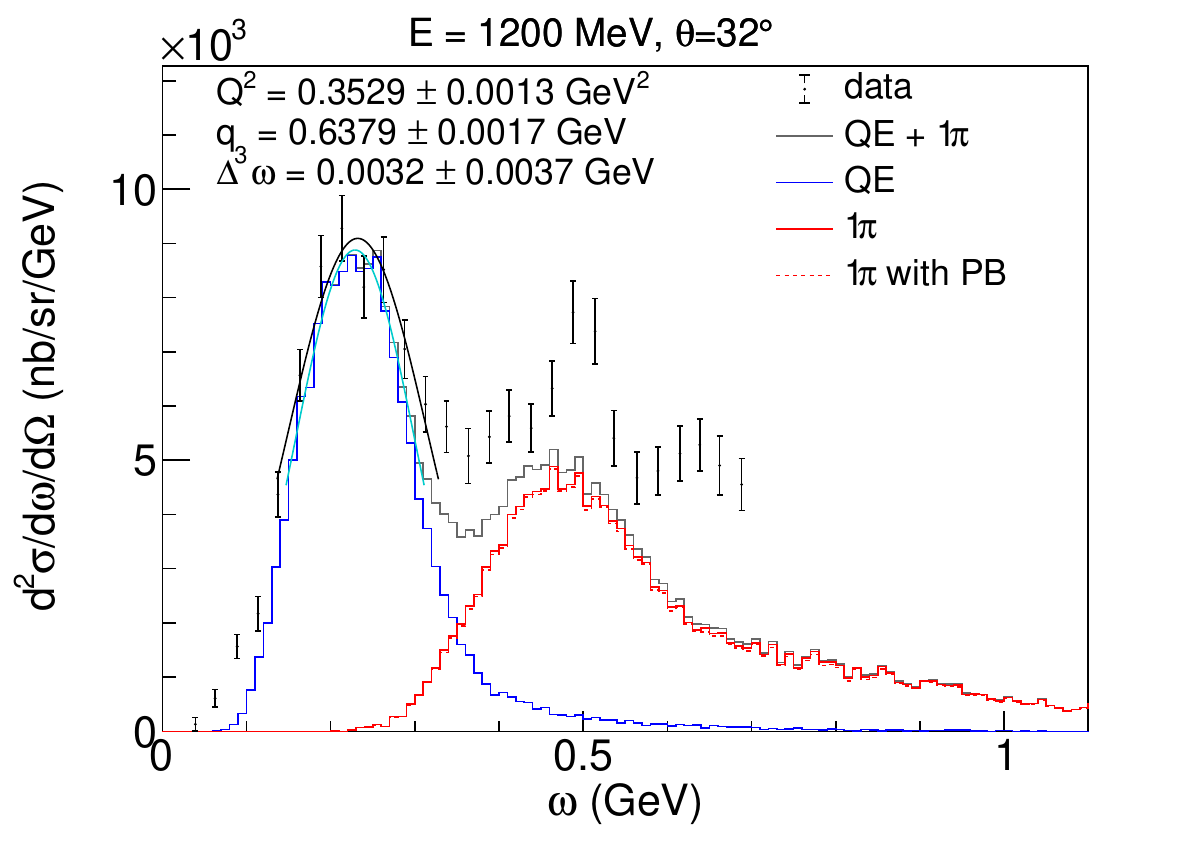}
\end{minipage}
\begin{minipage}[t]{0.30\textwidth} \centering
\includegraphics[width=1.0\textwidth]{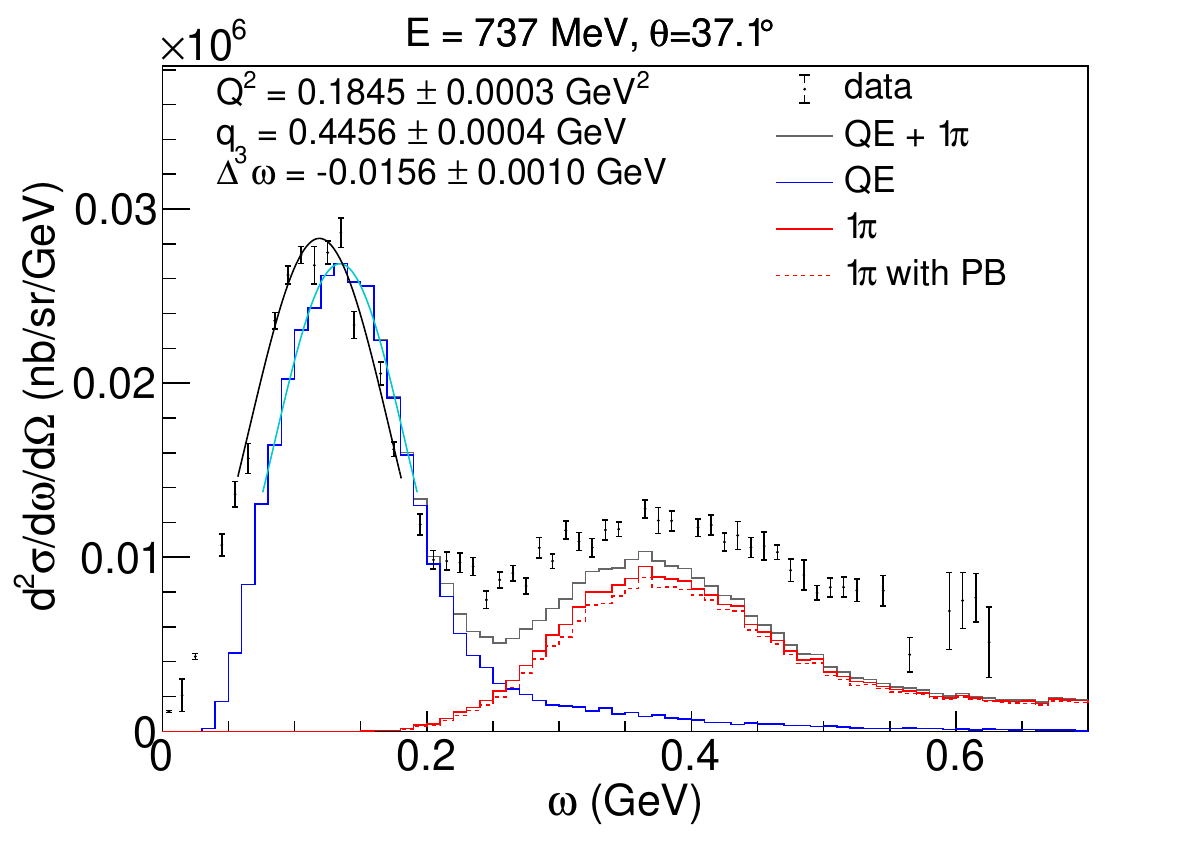}
\end{minipage}
\caption{
  The same as Fig.~\ref{fig:Barreau} but comparisons of inclusive electron scattering $^{16}{\rm O}(e,e')$ cross section between \textsc{NEUT} and experimental data~\cite{Anghinolfi:1996vm,O'Connell:1987ag}.
}
\label{fig:O16}
\end{figure*}

The \textsc{NEUT} simulations are compared with various experimental data at different electron energies and scattering angles~\citeall.
Radiative corrections are applied to the data, while Coulomb corrections are not included.
The results of $^{12}{\rm C}(e,e')$ are shown in Figs.~\ref{fig:Barreau}--\ref{fig:Day}, and those of $^{16}{\rm O}(e,e')$ are shown in Fig.~\ref{fig:O16}.
The simulations are performed considering the Coulomb potential and using the data-driven deexcitation model described in Sec~\ref{sec:formalism_qe}.
The choice of deexcitation model does not affect the results of the inclusive scattering discussed here.
Two peaks corresponding to QE and $1\pi$ interactions are observed.
Furthermore, it is observed that the \textsc{NEUT} simulation does not fully reproduce the QE peak in the datasets of low-momentum transfer.
This discrepancy is attributed to an effect beyond the PWIA, namely the distortion of the outgoing nucleon wave function by the nuclear potential~\cite{Bodek2019,PhysRevD.91.033005}.
More details are discussed in Sec.~\ref{sec:q3_eb}.
\par
In the region between QE and $1\pi$, known as the ``dip'' region, the excess of data is visible.
The excess is generally explained by multinucleon interaction, which is not yet implemented in the \textsc{NEUT} electron scattering.
The Valencia model by Nieves {\it et al.}~\cite{PhysRevD.88.113007} is implemented in \textsc{NEUT}, but it is limited to CC multinucleon interaction.
The solid (dashed) red line in Figs.~\ref{fig:Barreau}--\ref{fig:O16} shows the $1\pi$ contribution without (with) Pauli blocking.
This impact is small in the high-momentum transfer datasets where the $1\pi$ contribution is large.
The subsequent analysis is based on results without Pauli blocking for $1\pi$ interactions.
Although \textsc{NEUT} assumes that the target nucleon has on-shell mass, i.e., neglecting removal energy, there is overall good agreement with experimental data in the peak positions of $1\pi$.
Considering removal energy in $1\pi$ interactions is one of the important tasks for the future.
One possible approach is to adopt the de Forest approximation, as the $1\pi$ interaction introduced in Ref.~\cite{PhysRevC.100.045503}.
Achieving this would enable a more comprehensive evaluation of the \textsc{NEUT} model using the $1\pi$ region in electron scattering.

%%%%%%%%%%%%%%%%%%%%%%%%%%%%%%%

%%%%%%%%%%%%%%%%%%%%%%%%%%%%%%% O16

%%%
\subsection{Extracting momentum-dependent removal energy correction} \label{sec:q3_eb}
The QE peak shifts observed in Figs.~\ref{fig:Barreau}--\ref{fig:O16} show the limitation of the simple PWIA prescription, which neglects the distortion of the struck nucleon wave function by the nuclear potential.
These discrepancies can be addressed by applying a momentum-dependent removal energy correction extracted similarly to Ref.~\cite{Bodek2019,psf2023008005}.
The QE peaks in Figs.~\ref{fig:Barreau}--\ref{fig:O16} are fitted with a Gaussian function for both \textsc{NEUT} simulations and experimental data.
The peak shift $\Delta\omega$ is defined as 
\begin{equation} \label{eq:domega}
\Delta\omega = \omega_{data}-\omega_{neut},
\end{equation}
where $\omega_{data}\,(\omega_{neut})$ denotes the mean value of the Gaussian fitting of data (\textsc{NEUT}).
The Gaussian fit is performed multiple times, adjusting the fit range based on the mean and deviation obtained in each iteration, to mitigate bias from the tail.
The three-momentum transfer $q_{3}$ can be calculated with
\begin{equation}\begin{split} \label{eq:Q2_q3}
Q^2 &= 2E(E-\omega_{data})(1-\cos\theta),\\
q_{3} & = \sqrt{Q^2 + \omega_{data}^2},
\end{split} \end{equation}
where $E$ and $\theta$ denote the initial electron energy and scattering angle, respectively.
The following four datasets are excluded from this analysis due to the absence of clear peaks:
(200\,MeV,\,36$^\circ$),\,(200\,MeV,\,60$^\circ$),\,(240\,MeV,\,36$^\circ$),\,and (320\,MeV,\,36$^\circ$) on carbon.
\par
\begin{figure}[htbp] \centering
\includegraphics[width=1.0\columnwidth]{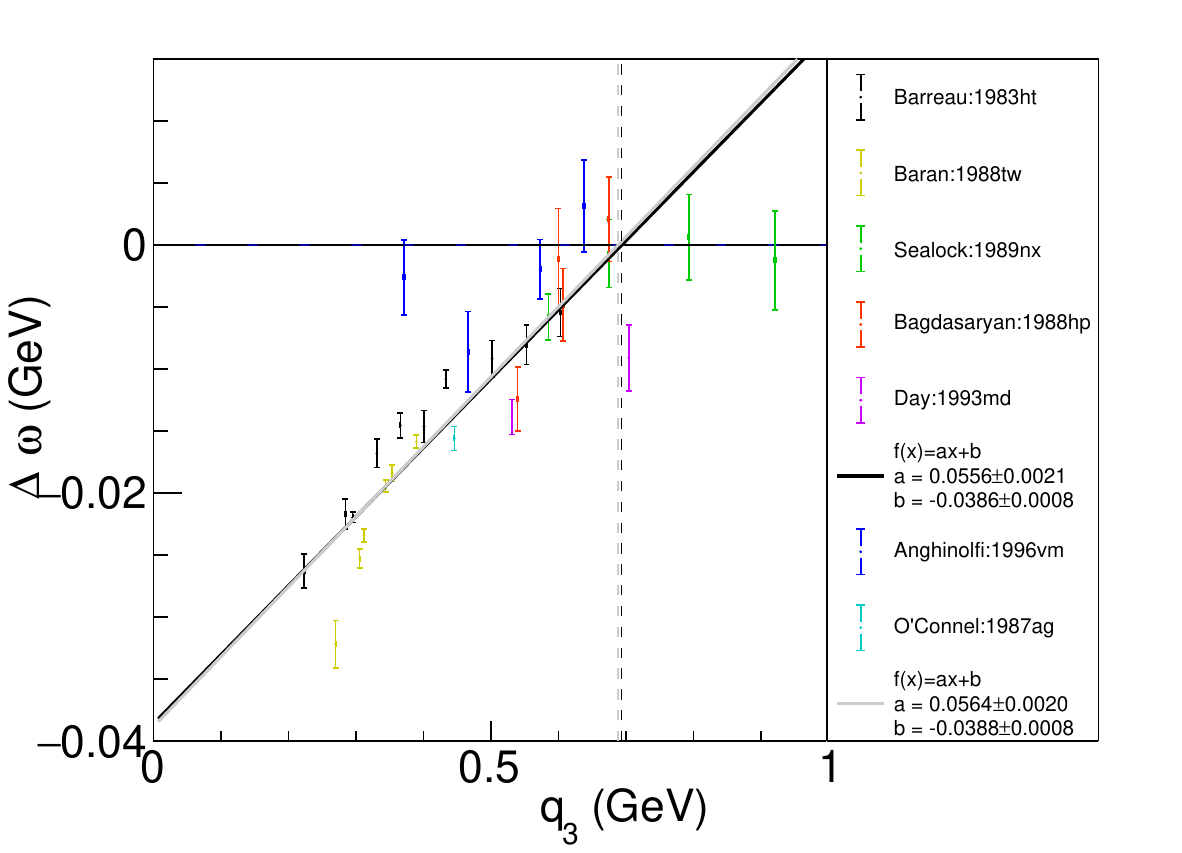}
\caption{
Relation between the peak shift $\Delta\omega$ and three-momentum transfer $q_3$.
The dots represent the results of 31 datasets in total: the blue and cyan dots represent data on oxygen obtained from Fig.~\ref{fig:O16}, while others represent data on carbon from Figs.~\ref{fig:Barreau}--\ref{fig:Day}.
The solid black line represents the best fit by a linear function $f(q_3)$ with carbon data only, 
while the solid gray line represents that using all data of carbon and oxygen.
The error bars represent the fitting uncertainty of Gaussian, mainly from the uncertainty of the experimental data.
}
\label{fig:dE_q3}
\end{figure}

Figure~\ref{fig:dE_q3} shows an obtained relation between the peak shift $\Delta\omega$ and the three-momentum transfer $q_3$.
A clear positive correlation is observed between these two parameters.
The peak shift deviates from zero and has a negative value in low-momentum transfer, while it is saturated to zero in high momentum transfer of $q_3\gtrsim0.7$\,GeV.
A similar trend is observed in numerical calculations based on the relativistic Fermi gas model~\cite{Bodek2019}.
This correlation is parametrized by fitting with a simple linear function $f(q_3)=aq_3+b$.
The following result is obtained when only the data on carbon is used:
\begin{equation}\begin{split} \label{eq:correction}
a &= 0.0556 \pm 0.0021, \\
b &= -0.0386 \pm 0.0008 {\rm\,GeV},
\end{split} \end{equation}
which is the black solid line in Fig.~\ref{fig:dE_q3}.
The errors denote the fitting uncertainties, including statistical uncertainty in \textsc{NEUT} and experimental uncertainty.
The data for oxygen have relatively large uncertainties, with no clear differences from carbon. 
Significant differences appear in the correlation of these parameters for heavier nuclei such as $^{40}$Ca,\,$^{56}$Fe, and $^{208}$Pb, but no significant differences are seen for carbon and oxygen~\cite{Bodek2019}.
When all data on carbon and oxygen are used, no significant change appears in the best-fit parameters as shown in Table~\ref{tab:corr_summary}.
\par
The derived function represents a correction to the removal energy, accounting for effects beyond the PWIA.
It reduces the effective removal energy at lower-momentum transfer.
We can empirically introduce effects beyond the PWIA by correcting the removal energy using this function.
More details are discussed in Sec.~\ref{sec:correction}.

\begin{table*}[htbp] \centering
\caption{
The best-fit parameters of momentum-dependent removal energy correction, $f(q_3)=aq_3+b$, with four analysis conditions.
The results are written with different treatment of Fermi momentum $p_F$, meson-exchange current contribution, and $^{16}$O data.
}
\label{tab:corr_summary}
\begin{tabular*}{0.70\textwidth}{@{\extracolsep{\fill}}lcccc} \hline \hline
Fermi momentum & MEC & $^{16}$O data & a & b (GeV) \\ \hline
$p_F=209\,$MeV & None & None & $0.0556 \pm 0.0021$ & $-0.0386 \pm 0.0008$ \\
               & None & Yes  & $0.0564 \pm 0.0020$ & $-0.0388 \pm 0.0008$ \\
               & Yes  & None & $0.0452 \pm 0.0034$ & $-0.0370 \pm 0.0013$ \\
               %& None & Yes & $0.0548 \pm 0.0021$ & $-0.0383 \pm 0.0008$ \\
               %& Yes & None & $0.0436 \pm 0.0035$ & $-0.0365 \pm 0.0013$ \\
$p_F=221\,$MeV & None & None & $0.0576 \pm 0.0021$ & $-0.0395 \pm 0.0008$ \\
\hline \hline
\end{tabular*}
\end{table*}

\subsubsection{Systematic uncertainties} \label{sec:q3_eb_syst}
The correction function derived above assumes a Fermi momentum of 209\,MeV, and neglects multinucleon interaction and deep inelastic scattering (DIS).
The effects of the Coulomb potential and Pauli blocking on the determination of the correction function are negligibly small as discussed in Appendices~\ref{sec:appen_coul} and \ref{sec:appen_pb}.
Varying the Fermi surface to $p_F=221$\,MeV, which is derived from electron scattering data based on Fermi gas model~\cite{PhysRevLett.26.445}, results in a minor change comparable to the uncertainty, as shown in Table~\ref{tab:corr_summary}.
The DIS contribution to the QE peak is also small, as inferred from Ref.~\cite{PhysRevD.103.113003}.
However, multinucleon interaction contributes significantly to the determination of the QE peak position as shown in Refs.~\cite{PhysRevD.103.113003,PhysRevC.100.045503}.
Therefore, the result above, which neglects multinucleon interactions, may have some bias.
An accurate assessment of this contribution requires the implementation of multinucleon interaction into \textsc{NEUT}, which is one of the significant future challenges.
Instead of implementation, the impact of multinucleon interactions is investigated by using numerical calculations on carbon by Rocco {\it et al.}~\cite{PhysRevC.100.045503}, which considers meson-exchange current (MEC).
\par
\begin{figure*}[htbp] \centering
\begin{minipage}[t]{0.32\textwidth} \centering
\includegraphics[width=1.0\textwidth]{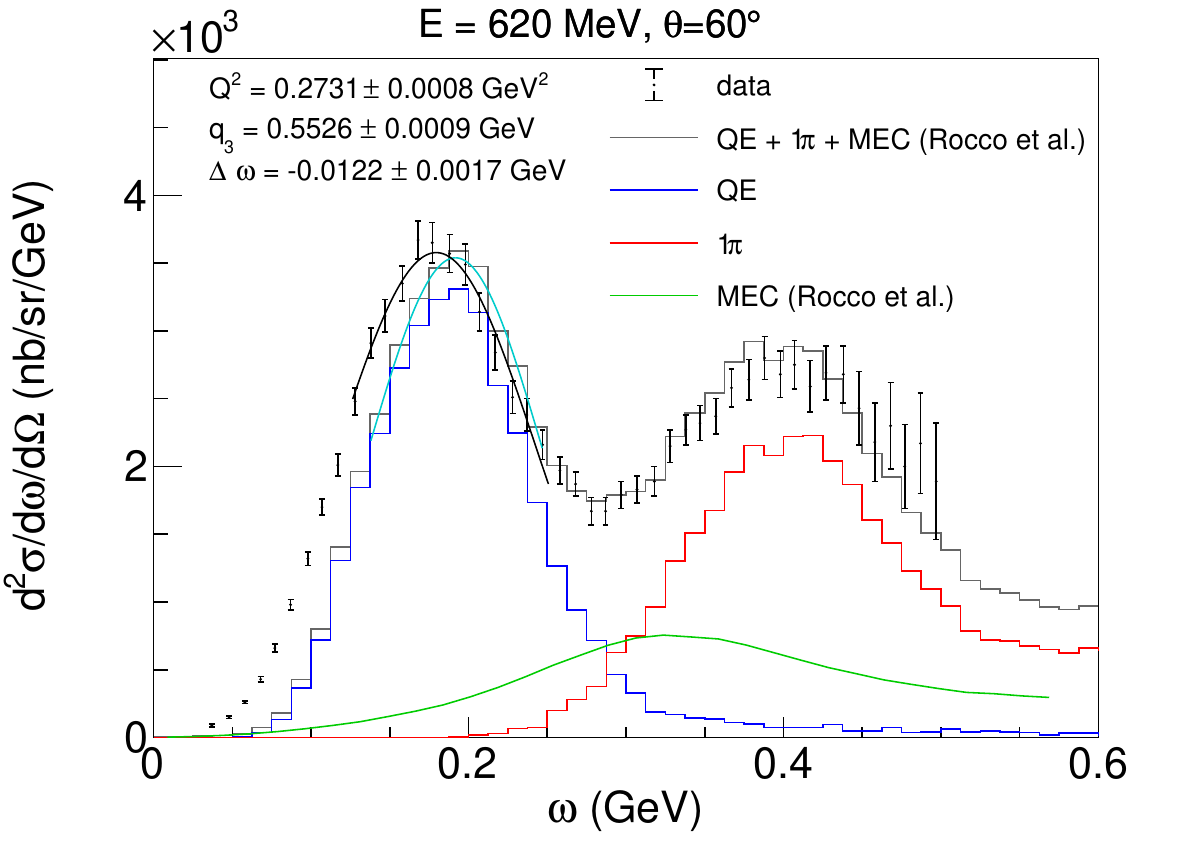}
\end{minipage}
\begin{minipage}[t]{0.32\textwidth} \centering
\includegraphics[width=1.0\textwidth]{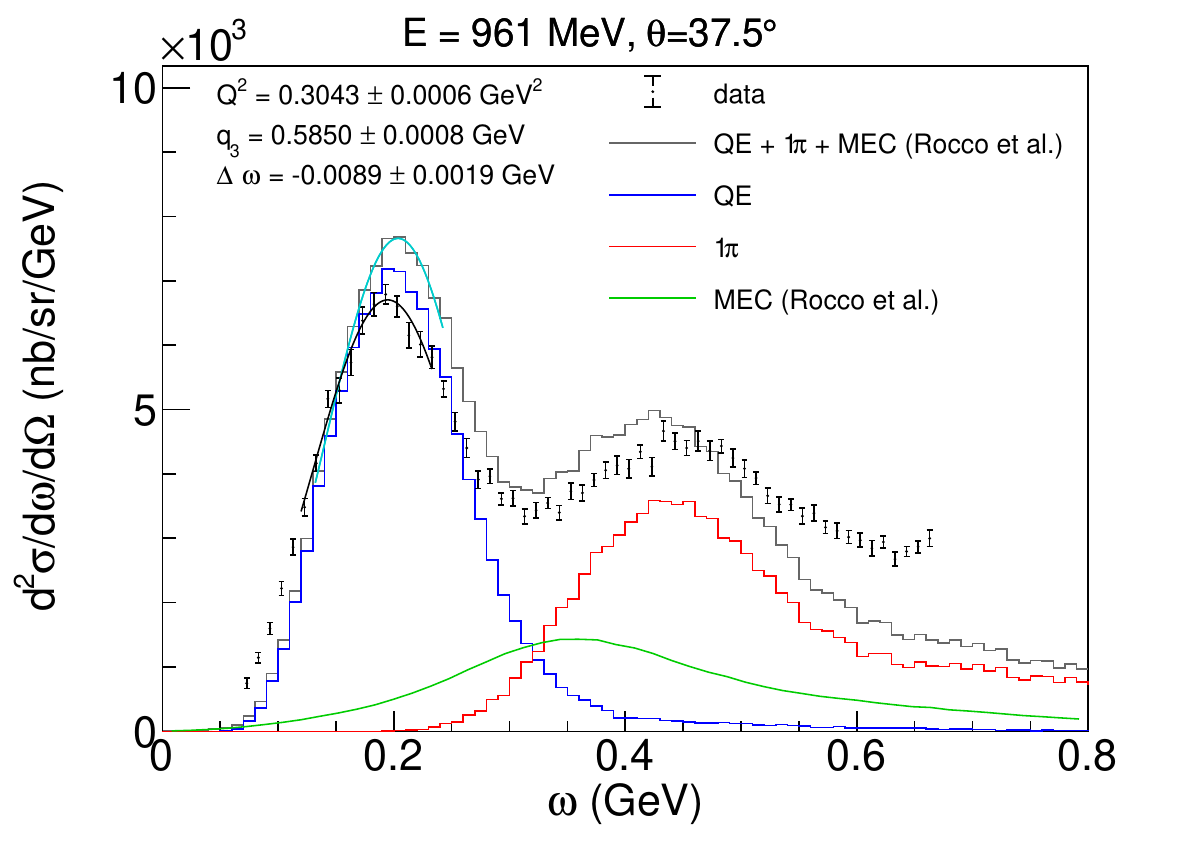}
\end{minipage}
\begin{minipage}[t]{0.32\textwidth} \centering
\includegraphics[width=1.0\textwidth]{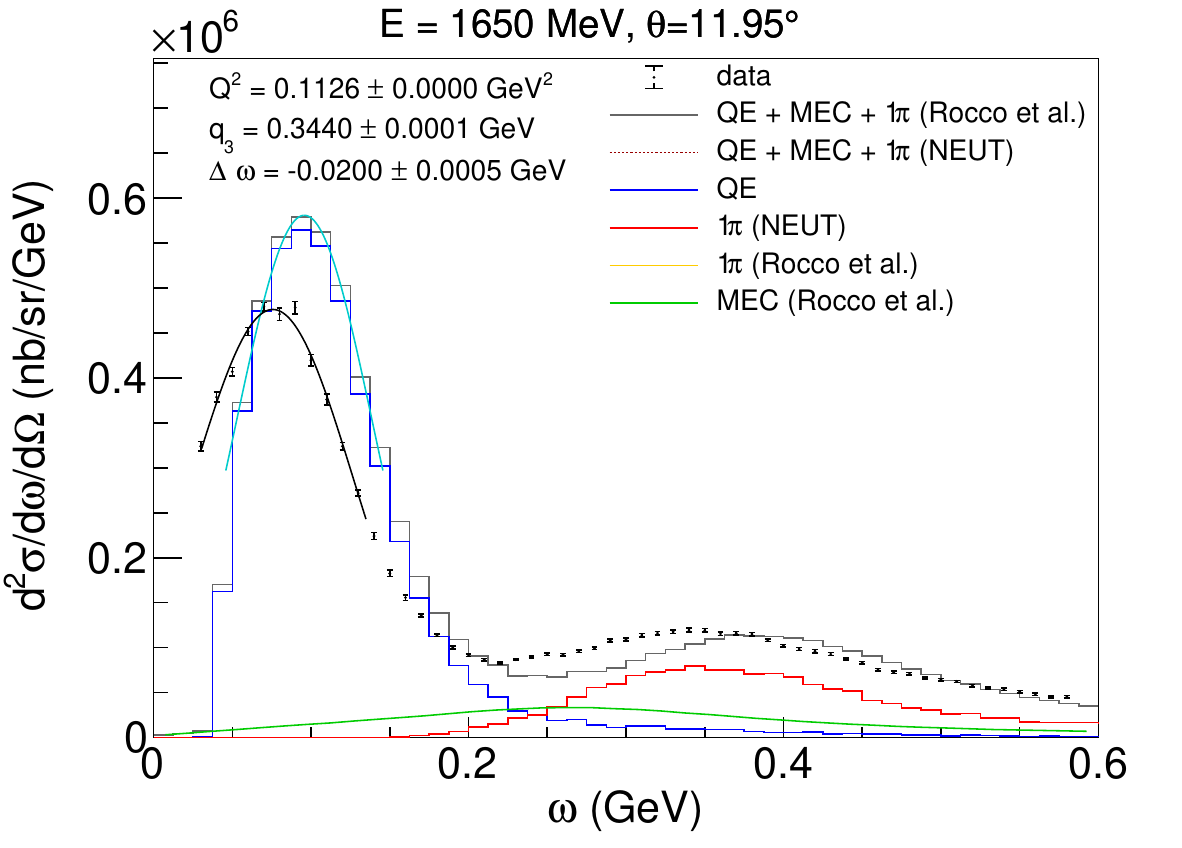}
\end{minipage}
\caption{Comparisons of inclusive electron scattering $^{12}$C$(e,e')$ cross section between \textsc{NEUT}, numerical calculation of MEC by Rocco {\it et al.}~\cite{PhysRevC.100.045503}, and experimental data~\cite{Barreau:1983ht,Baran:1988tw,Sealock:1989nx}.
The gray line shows the sum of QE and $1\pi$ from \textsc{NEUT}, and MEC by Rocco {\it et al.}.
}
\label{fig:MEC}
\end{figure*}

Figure~\ref{fig:MEC} shows comparisons between NEUT simulations and experimental data, this time including the MEC contribution~\cite{PhysRevC.100.045503}.
Including the MEC contribution improves the agreement between \textsc{NEUT} and the experimental data, particularly in the dip region.
Also, by comparing the QE peak positions of Fig.~\ref{fig:MEC} and Figs.~\ref{fig:Barreau}--\ref{fig:Sealock}, we can see the impact of MEC on the QE peak position of \textsc{NEUT}.
By adding MEC contribution, the peak positions of \textsc{NEUT} shift toward higher energy transfer by several MeV.
This shift is larger than that caused by other systematic factors, such as Pauli blocking. 
Consequently, the absence of MEC likely contributes as a primary systematic uncertainty in the analysis above.
As inferred from the figures, the magnitude of moving depends on the datasets, i.e., momentum transfer.

\begin{figure}[htbp] \centering
\includegraphics[width=1.0\columnwidth]{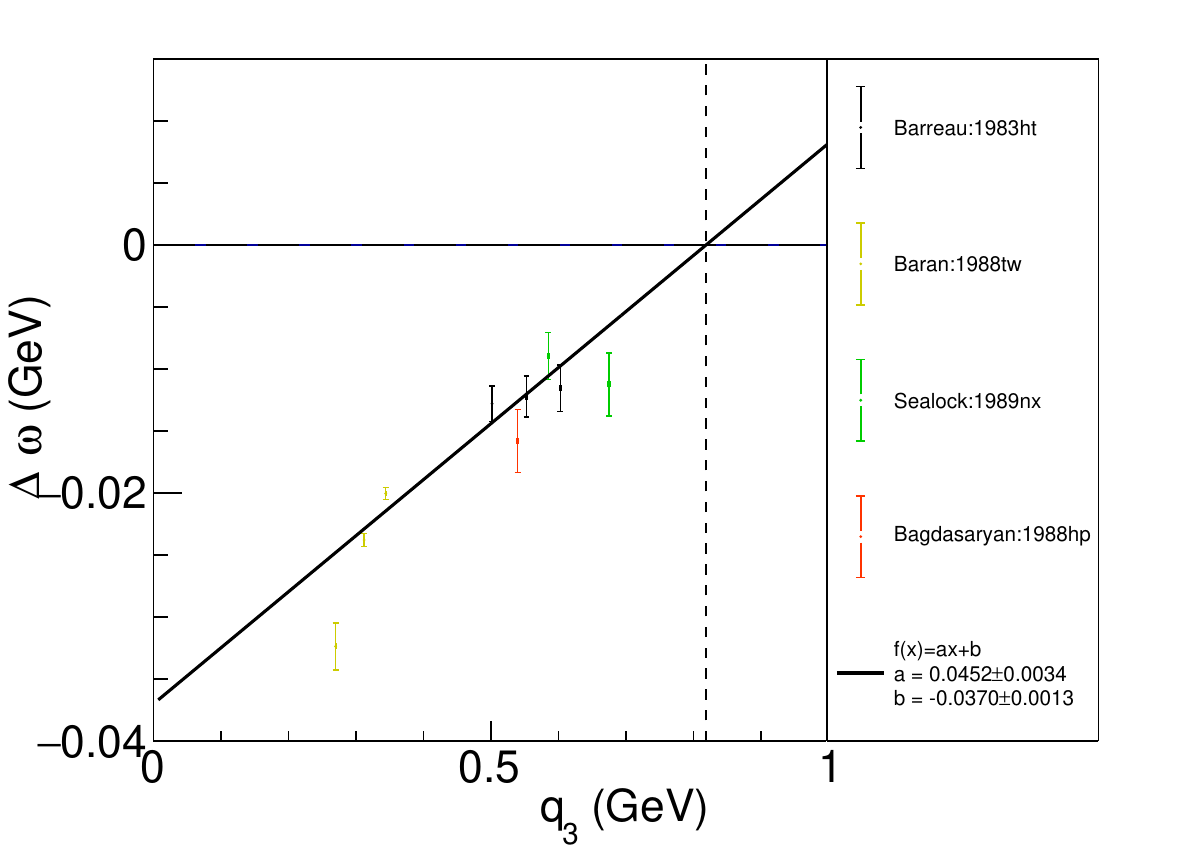}
\caption{
The same as Fig.~\ref{fig:dE_q3}, but meson-exchange current contribution calculated by Rocco {\it et al.}~\cite{PhysRevC.100.045503} is added.
}
\label{fig:dE_q3_2p2h}
\end{figure}

The correlation between the peak shift and three-momentum transfer is reevaluated, including the MEC contribution calculated by Rocco {\it et al.}
Figure~\ref{fig:dE_q3_2p2h} shows the results, noting that the MEC calculations are limited to certain datasets.
By fitting the correlation with a linear function, the following results are obtained (see also Table~\ref{tab:corr_summary}):
\begin{equation}\begin{split} \label{eq:correction_2p2h}
a &= 0.0452 \pm 0.0034,\\
b &= -0.0370 \pm 0.0013 {\rm\,GeV},
\end{split} \end{equation}
with larger uncertainties due to the limited datasets.
It is observed that the impact of MEC makes a major difference in determining correction function compared to Pauli blocking and Coulomb potential.
This result is used to assess the systematic uncertainty of the MEC contribution in the following discussion.

Another possible systematic is the removal energy of $1\pi$ interactions.
As mentioned in Sec.~\ref{sec:formalism_1pi}, \textsc{NEUT} is not able to consider removal energy for $1\pi$ interactions, which might also affect the QE peak position.
The effect of this assumption is investigated by using the numerical calculation by Rocco {\it et al.}~\cite{PhysRevC.100.045503} again.
Their calculation of $1\pi$ interactions is based on the DCC model considering the SF using the de Forest approximation.
We found differences in overall distribution; their calculations have larger energy transer.
The $1\pi$ tail component near the QE peak is not so significant, leading to minor effect on the QE peak position.
Therefore, the uncertainty from the removal energy in $1\pi$ is neglected in the following analysis.
\par
In this analysis, uncertainties are taken from the archived data release~\cite{benhar2006archivequasielasticelectronnucleusscattering}, and no additional sources of uncertainty, such as normalization, are incorporated.
The correlation between uncertainties across different datasets is not taken into account.
While these simplifications are made in this study, addressing this limitation in future work will be crucial.
Incorporating additional uncertainties and their correlations is expected to enhance the accuracy of the quantitative discussion, leading to more robust and reliable results.

%%%%%%%%%%%%%%%%%%%%%%%%%%%%%
\section{Applying momentum-dependent removal energy correction} \label{sec:correction}
The momentum-dependent removal energy correction, derived in Sec.~\ref{sec:q3_eb}, is applied to account for effects beyond PWIA empirically. 
The correction $\Delta \tilde{E}$, as a function of three-momentum transfer $q_3$, is defined as follows.
\begin{equation}\begin{split}
\Delta \tilde{E} & =  
\begin{dcases}
f(q_3) & (f(q_3)\leq0),\\
0 & (f(q_3)>0),
\end{dcases}
\end{split}\end{equation}
where $f(q_3)$ is the best-fit function shown in Eq.~(\ref{eq:correction}), which is extracted from carbon data only.
Since no clear differences are observed between carbon and oxygen, this best-fit function is applied to both carbon and oxygen.
Above a certain three-momentum transfer $q_3$, the function $f(q_3)$ becomes positive, and $\Delta\tilde{E}$ is truncated to zero.
That is, no correction is applied at high-momentum transfers.
This truncation reflects the fact that the peak shift is not observed in high-momentum transfer datasets, i.e., the PWIA is sufficiently accurate in this region.
The differential cross section is then calculated by redefining the removal energy, shown in Eq.~(\ref{eq:etilde}), as 
\begin{equation}\begin{split}
\tilde{E} \rightarrow \tilde{E} + \Delta\tilde{E}.
\end{split}\end{equation}
This correction reduces the removal energy since $\Delta\tilde{E}$ has a negative value, which could potentially lead to an unphysical negative removal energy.
A truncation is introduced to maintain physical consistency:
\begin{equation}\begin{split}\label{eq:trunc_removal}
\tilde{E} =
\begin{dcases}
0 & (\tilde{E} < 0), \\
\tilde{E} & (\tilde{E}\geq0),
\end{dcases}
\end{split}\end{equation}
where the negative removal energy is not allowed.
In low-momentum transfer datasets with large corrections, this truncation occurs frequently and reduces the correction effect.
While requiring positive removal energy is a natural assumption, the following analysis evaluates both results with and without this truncation.

%%%
\subsection{Electron scattering} \label{sec:qe_eb_apply_electron}
\begin{figure*}[htbp] \centering
\begin{minipage}[b]{0.495\textwidth} \centering
\includegraphics[width=1.0\textwidth]{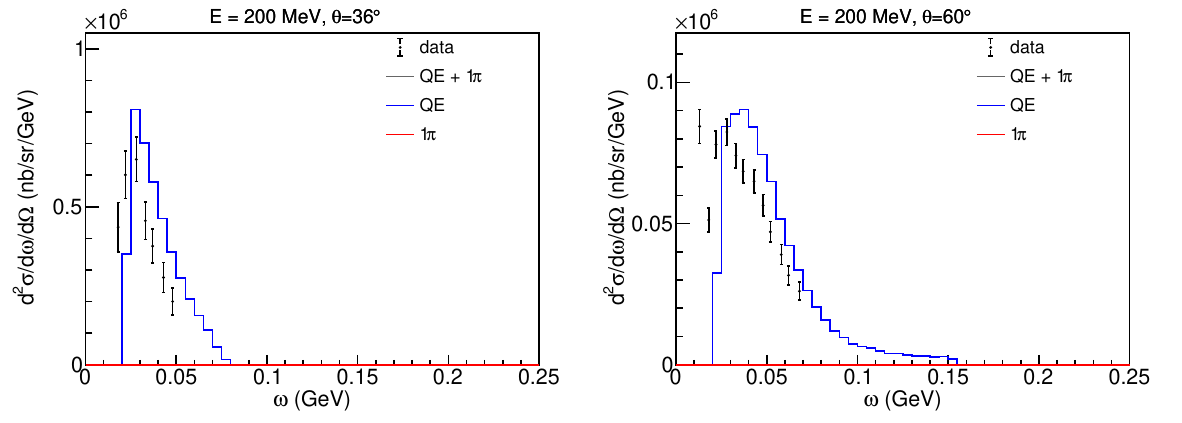}
\end{minipage}
\begin{minipage}[b]{0.495\textwidth} \centering
\includegraphics[width=1.0\textwidth]{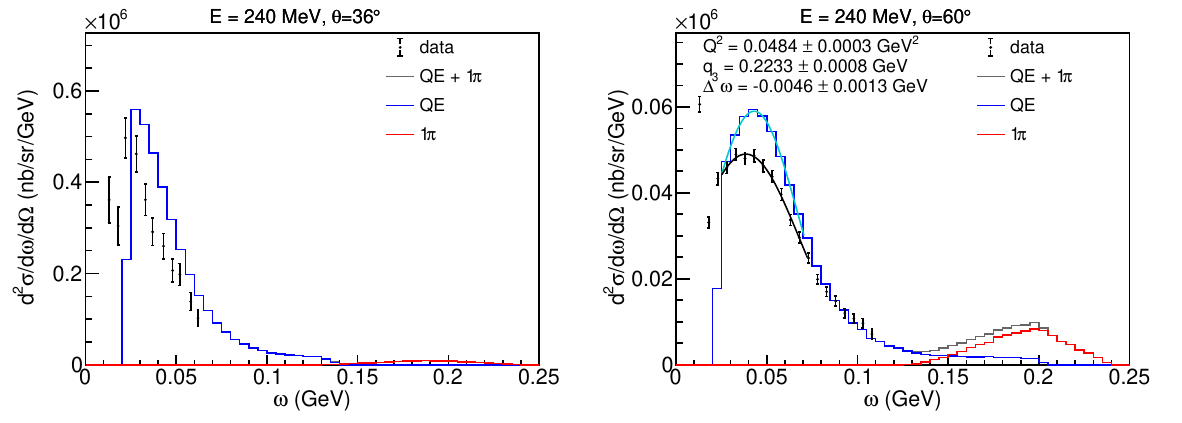}
\end{minipage}
\\
\begin{minipage}[b]{0.495\textwidth} \centering
\includegraphics[width=1.0\textwidth]{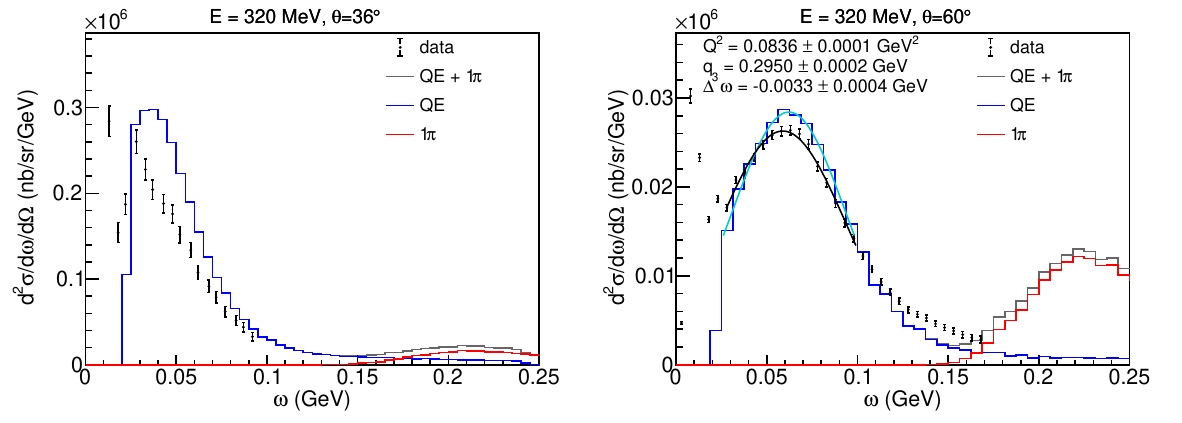}
\end{minipage}
\begin{minipage}[b]{0.495\textwidth} \centering
\includegraphics[width=1.0\textwidth]{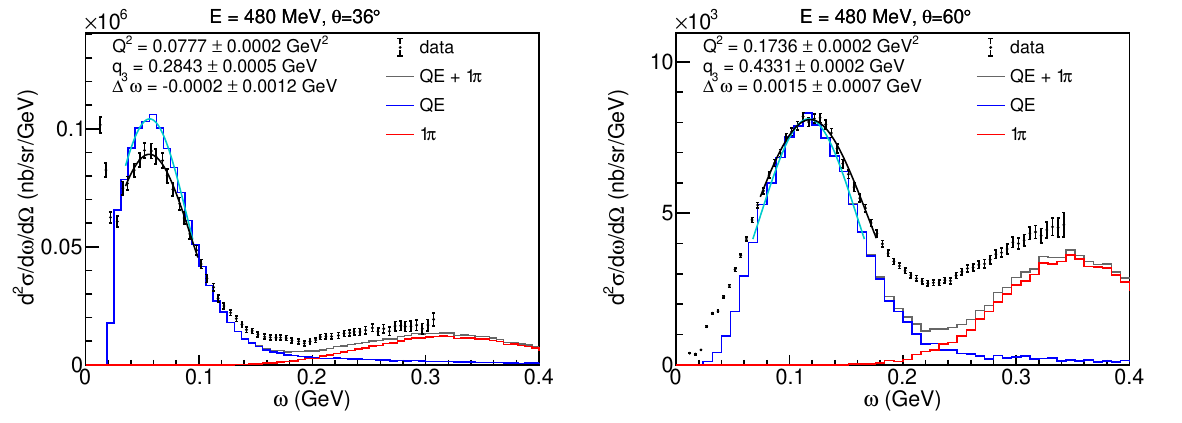}
\end{minipage}
\caption{
The same as Fig.~\ref{fig:Barreau} but after applying the momentum-dependent removal energy correction of Eq.~(\ref{eq:correction}), where negative removal energy is not allowed.
The figures that do not show the Gaussian fit (black solid line) correspond to the datasets excluded from the analysis used to obtain the correction in Sec.~\ref{sec:q3_eb}.
}
\label{fig:Barreau_corr}
\end{figure*}

\begin{figure*}[htbp] \centering
\begin{minipage}[b]{0.495\textwidth} \centering
\includegraphics[width=1.0\textwidth]{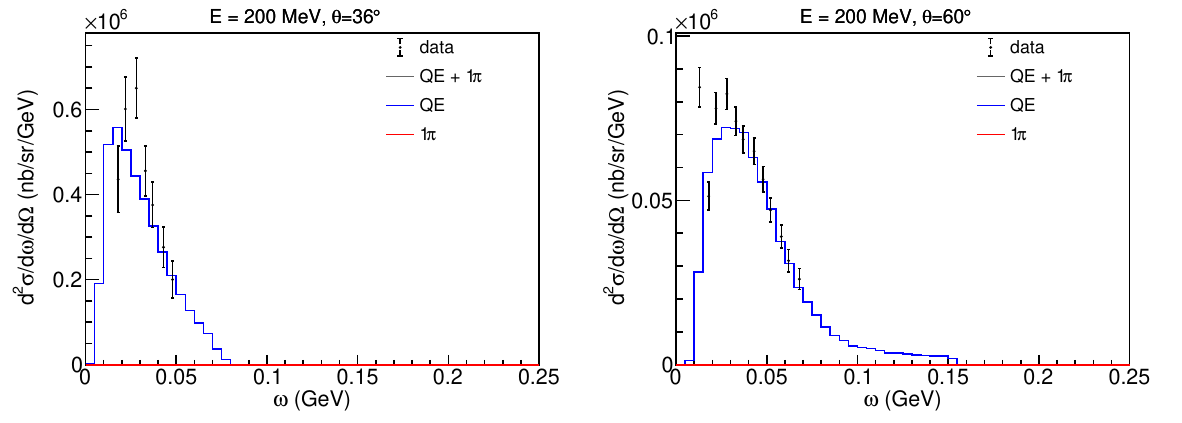}
\end{minipage}
\begin{minipage}[b]{0.495\textwidth} \centering
\includegraphics[width=1.0\textwidth]{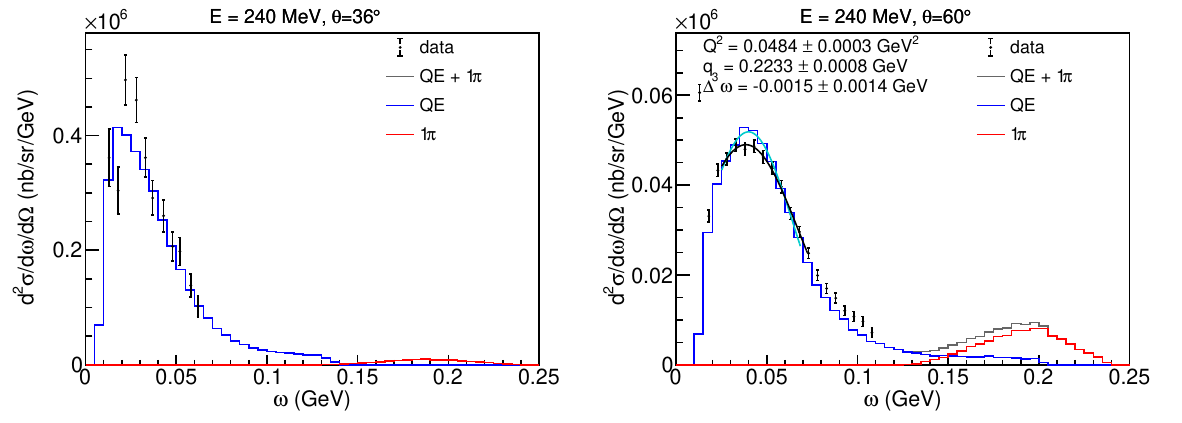}
\end{minipage}
\\
\begin{minipage}[b]{0.495\textwidth} \centering
\includegraphics[width=1.0\textwidth]{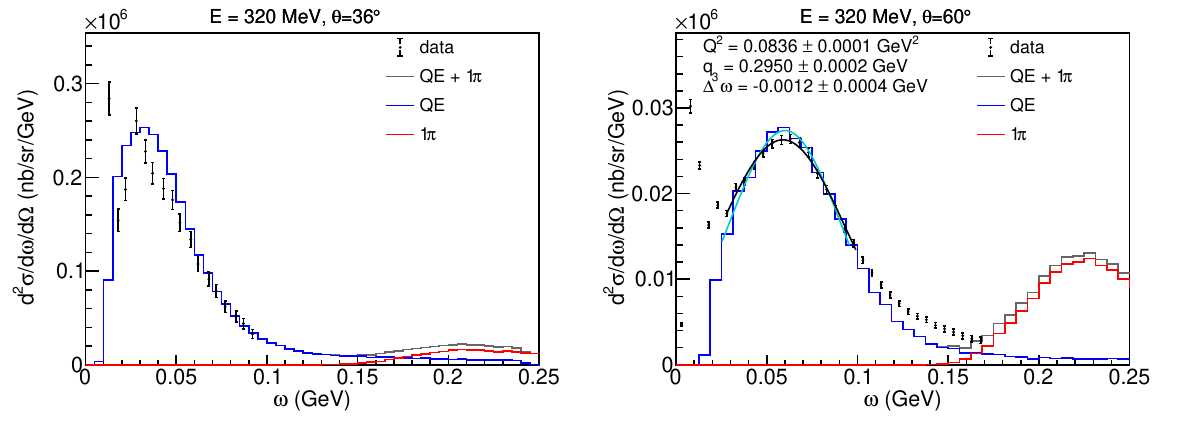}
\end{minipage}
\begin{minipage}[b]{0.495\textwidth} \centering
\includegraphics[width=1.0\textwidth]{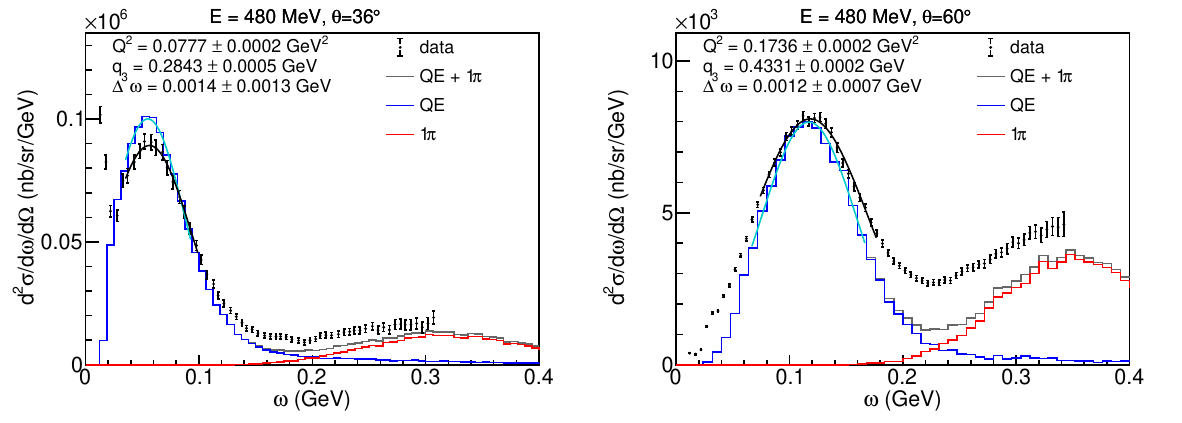}
\end{minipage}
\caption{
The same as Fig.~\ref{fig:Barreau} but after applying the momentum-dependent removal energy correction of Eq.~(\ref{eq:correction}), where negative removal energy is allowed.
The figures that do not show the Gaussian fit (black solid line) correspond to the datasets excluded from the analysis used to obtain the correction in Sec.~\ref{sec:q3_eb}.
}
\label{fig:Barreau_corr_trunc}
\end{figure*}

The correction is first applied to electron scattering in order to investigate its validity and effectiveness.
Figures~\ref{fig:Barreau_corr} and \ref{fig:Barreau_corr_trunc} show the results after applying the correction.
Figure.~\ref{fig:Barreau_corr_trunc} shows the result without the truncation as defined in Eq.~(\ref{eq:trunc_removal}), allowing negative removal energy.
These figures include datasets excluded from the analysis used to obtain the correction in Sec.~\ref{sec:q3_eb}, to check the validity of the correction's extrapolation.
By applying the correction, the QE cross section of \textsc{NEUT} is shifted and the relative difference from the experimental data is mitigated.
If positive removal energy is required (Fig.~\ref{fig:Barreau_corr}), the discrepancy still remains in low-momentum transfer datasets because the correction effect is reduced.
This indicates that the extrapolation of the correction to these datasets was unsuccessful.
On the other hand, if negative removal energy is accepted (Fig.~\ref{fig:Barreau_corr_trunc}), a better agreement is observed.
While a peak shape discrepancy is still present, no significant peak position difference is observed.
Although allowing negative removal energy is unphysical, this approach demonstrates, to some extent, the validity of the correction's extrapolation to these datasets.
\par
The correction discussed here is empirical and has limitations in its ability to reproduce the experimental data with better accurary.
Moreover, it is challenging to obtain good agreement with experimental data while maintaining physical consistency using such an empirical method.
A theory-driven correction discussed in Ref.~\cite{PhysRevD.91.033005} shows better agreement with the experimental data in both peak position and shape, compared to this result.
This is because they account for the shift of the QE peak position using the real part of the optical potential, and a broadening of the cross section induced by FSI.
The inclusion of such a theory-driven correction is one of the essential challenges for future work.

%%%
\subsection{Neutrino interactions} \label{sec:qe_eb_apply_neutrino}
The correction is applied to neutrino scattering to evaluate its impact on neutrino energy reconstruction in neutrino oscillation experiments.
The neutrino energy is reconstructed from the measured charged lepton kinematics assuming CCQE interactions, as described in the T2K far detector analysis~\cite{Abe2023,PhysRevD.96.092006}.
The neutrino energy in CCQE interactions is reconstructed using energy and momentum conservation:
\begin{equation}\begin{split}
 E_{\nu}^{rec} &= \frac{2 E_l \tilde{M} - (m_l^2 + \tilde{M}^2  - M_f^2)}{2(\tilde{M} - E_l + p_l \cos\theta_l)}, \\
 \tilde{M} &= M_i - E_b,
\end{split} \end{equation}
where $E_l$, $p_l$, $m_l$, and $\cos\theta_l$ denote outgoing charged lepton energy, momentum, mass, and scattering angle, respectively.
$M_i$ and $M_f$ are the initial and final nucleon masses, and $E_b$ is binding energy.
For the following discussion, the same binding energy of $E_b=27$\,MeV for $^{16}$O, as assumed in T2K~\cite{PhysRevD.96.092006}, is used.
\par
\begin{figure}[htbp] \centering
\begin{minipage}[t]{0.82\columnwidth} \centering
\includegraphics[width=1.0\textwidth]{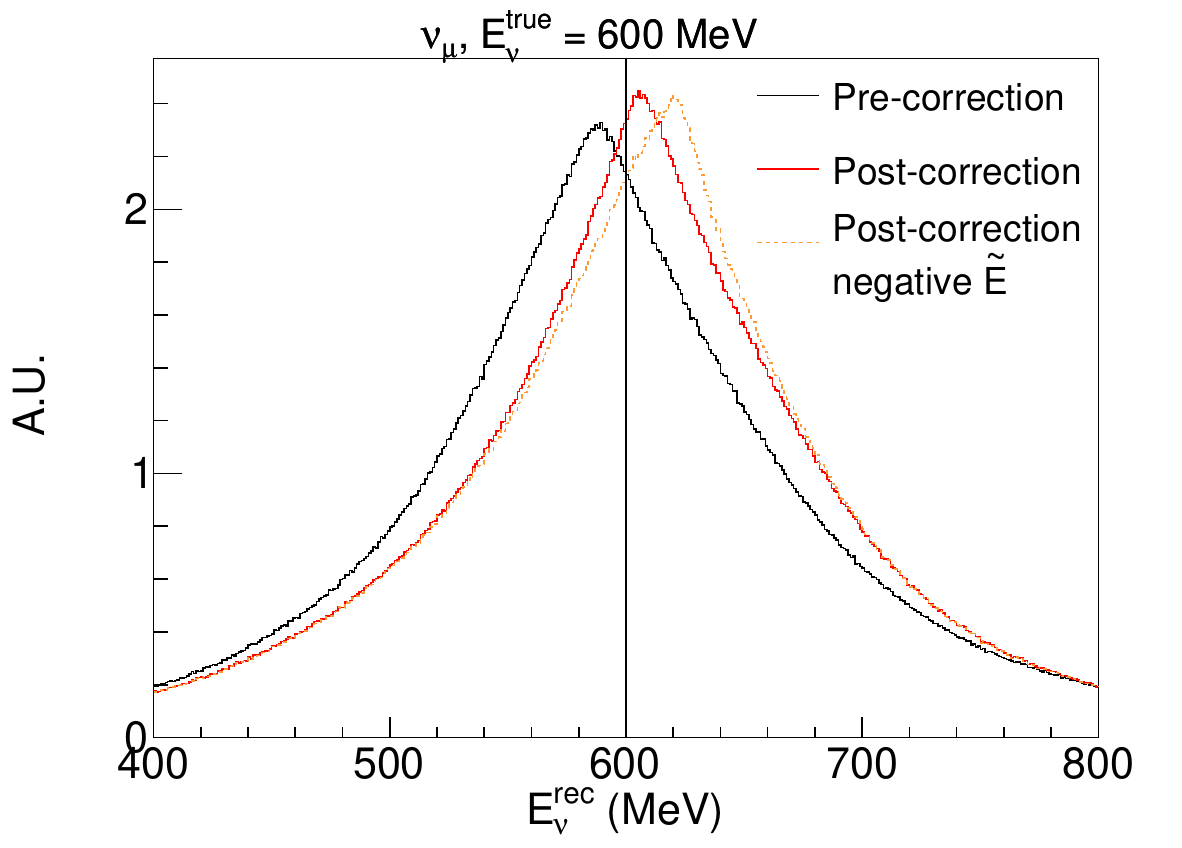}
\end{minipage}
\\ \vspace{5pt}
\begin{minipage}[t]{0.82\columnwidth} \centering
\includegraphics[width=1.0\textwidth]{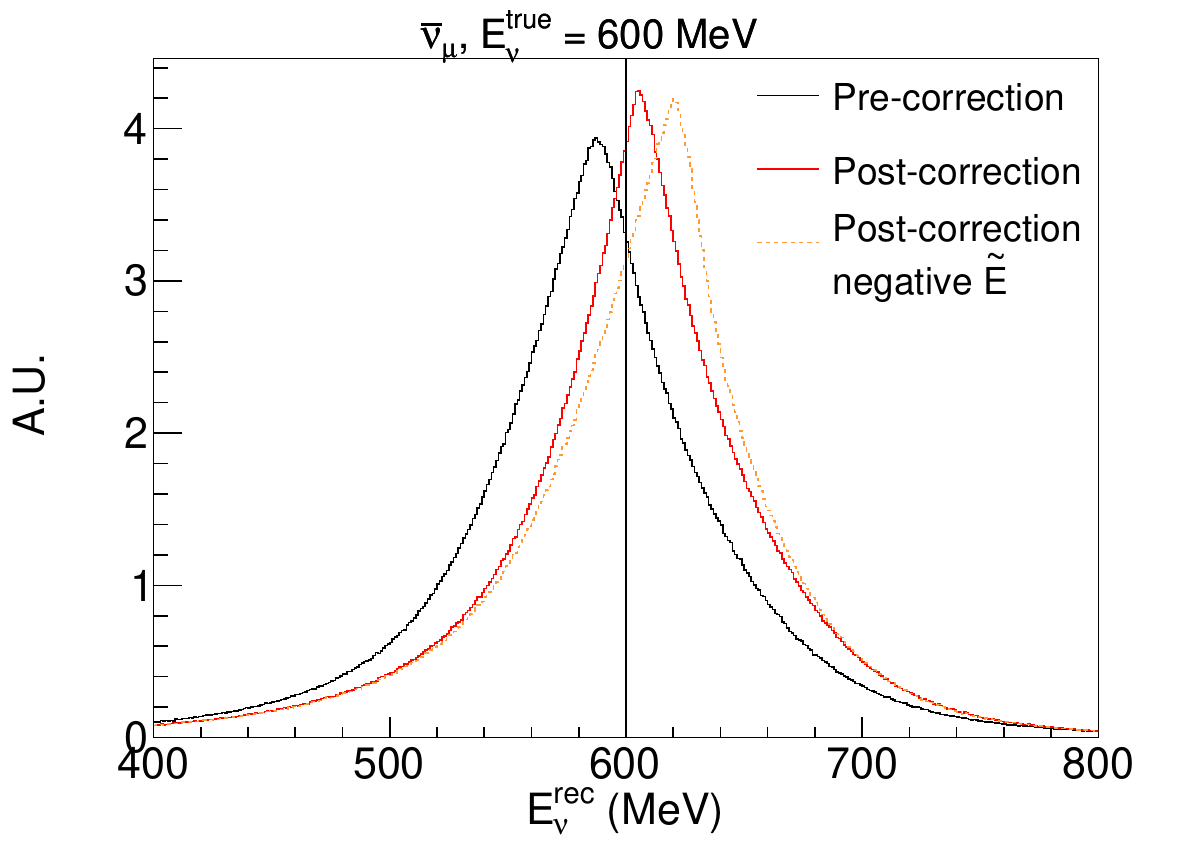}
\end{minipage}
\caption{
Reconstructed neutrino energy distributions in $\nu_\mu$ (top) and $\bar{\nu}_\mu$ (bottom) CCQE interactions on oxygen, at a true neutrino energy of 600 MeV.
The black line represents the calculations before applying the momentum-dependent removal energy correction.
Two results after the correction are shown: the red line corresponds to the case where negative removal energy is not allowed while the orange dashed line allows it.
}
\label{fig:O16_recon}
\end{figure}

Figure~\ref{fig:O16_recon} shows the reconstructed neutrino energy of $\nu_\mu$ and $\bar{\nu}_\mu$ CCQE interactions on oxygen at a true neutrino energy of 600\,MeV.
The correction derived from electron scattering results in a significantly different distribution.
A sizable change in peak positions is observed: approximately 20 MeV when positive removal energy is required and 30\,MeV when negative removal energy is allowed.
The peak positions of the distribution for different true neutrino energies and neutrino flavors are summarized in Table~\ref{tab:O16_recon}.
When negative removal energy is not allowed, the peak position shift does not show a clear dependence on the true neutrino energy or flavor.
On the other hand, when it is allowed, low-energy neutrinos have larger peak shift by about 2\,MeV.
This is because the large correction applied to events with low-momentum transfer is weakened by the truncation.
\par
The impact of the major systematic uncertainty, namely the MEC contribution, is further investigated using the best-fit function in Eq.~(\ref{eq:correction_2p2h}), which accounts for the MEC contribution.
While including MEC leads to a noticeable difference in the correction function, it results in only minor shifts, within $\pm1\,$MeV in the peak positions of the reconstructed neutrino energy.

\begin{table}[htbp] \centering
\caption{Peak positions of the reconstructed neutrino energy distribution before and after applying the removal energy correction for different true neutrino energies and flavors.
Two results after the correction are shown. 
The results denoted as (negative $\tilde{E}$) correspond to the results allowing negative removal energy.
All values are written in units of MeV.
}
\label{tab:O16_recon}
\begin{tabular*}{1.0\columnwidth}{@{\extracolsep{\fill}}llccccc} \hline \hline
 &  & \multicolumn{5}{c}{True neutrino energy}  \\
Flavor & Correction                      & 200 & 400 & 600 & 800 & 1000 \\ \hline
$\nu_\mu$ & before                       & 191 & 389 & 588 & 788 & 988 \\
          & after                        & 209 & 407 & 607 & 806 & 1006 \\
          & after (negative $\tilde{E}$) & 223 & 420 & 619 & 819 & 1018 \\
$\nu_e$ & before                       & 189 & 388 & 588 & 788 & 988 \\
        & after                        & 207 & 407 & 606 & 806 & 1006 \\
        & after (negative $\tilde{E}$) & 222 & 420 & 619 & 818 & 1018 \\
$\bar{\nu}_\mu$ & before                       & 189 & 388 & 588 & 788 & 988 \\ 
                & after                        & 207 & 406 & 606 & 806 & 1006 \\ 
                & after (negative $\tilde{E}$) & 222 & 420 & 620 & 819 & 1019 \\ 
$\bar{\nu}_e$ & before                       & 187 & 387 & 587 & 787 & 987 \\ 
              & after                        & 205 & 405 & 606 & 806 & 1006 \\ 
              & after (negative $\tilde{E}$) & 221 & 420 & 620 & 819 & 1019 \\ 
\hline \hline
\end{tabular*}
\end{table}

\begin{figure*}[htbp] \centering
\begin{minipage}[t]{0.32\textwidth} \centering
\includegraphics[width=1.0\textwidth]{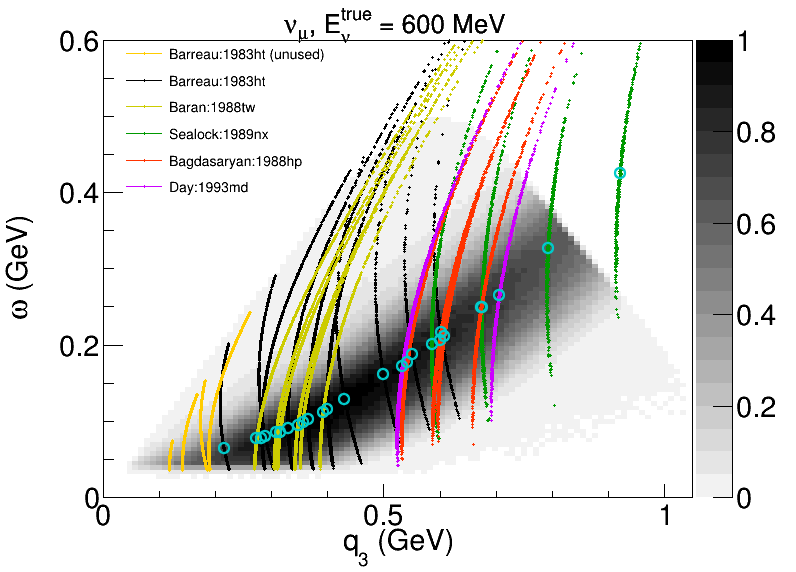}
\end{minipage}
\begin{minipage}[t]{0.32\textwidth} \centering
\includegraphics[width=1.0\textwidth]{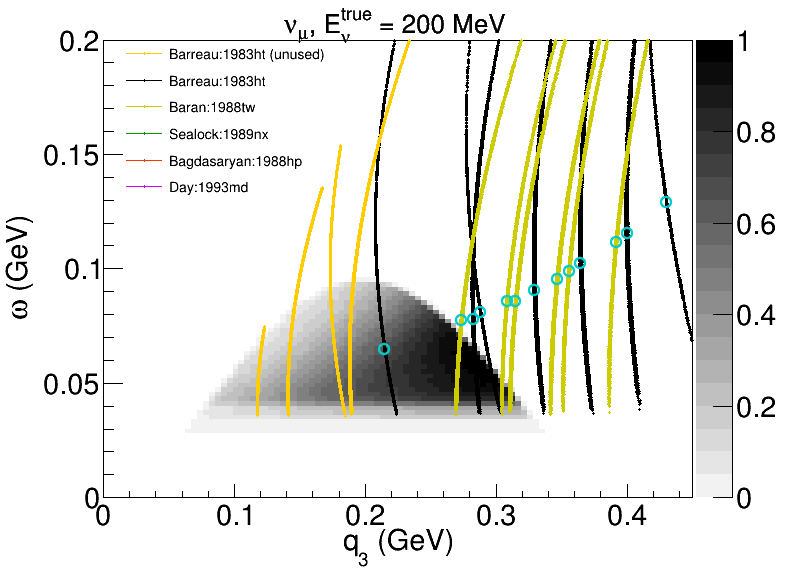}
\end{minipage}
\begin{minipage}[t]{0.32\textwidth} \centering
\includegraphics[width=1.0\textwidth]{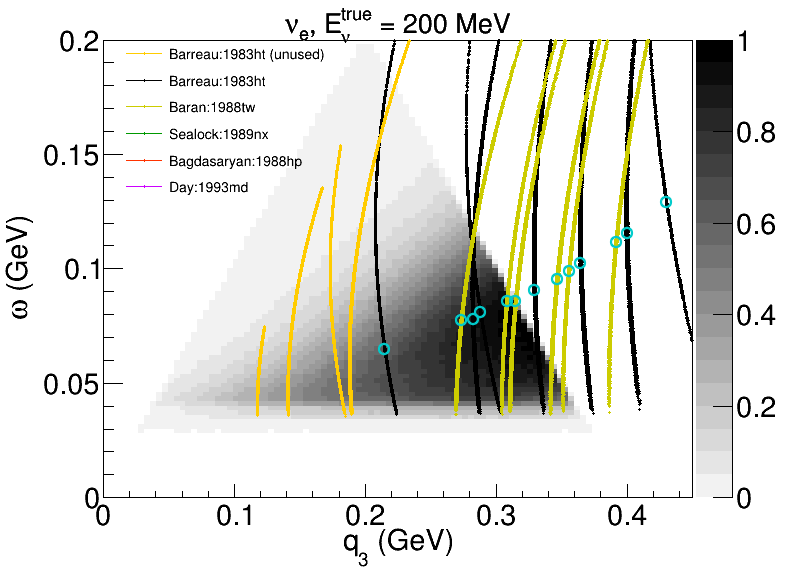}
\end{minipage}
\\
\begin{minipage}[t]{0.32\textwidth} \centering
\includegraphics[width=1.0\textwidth]{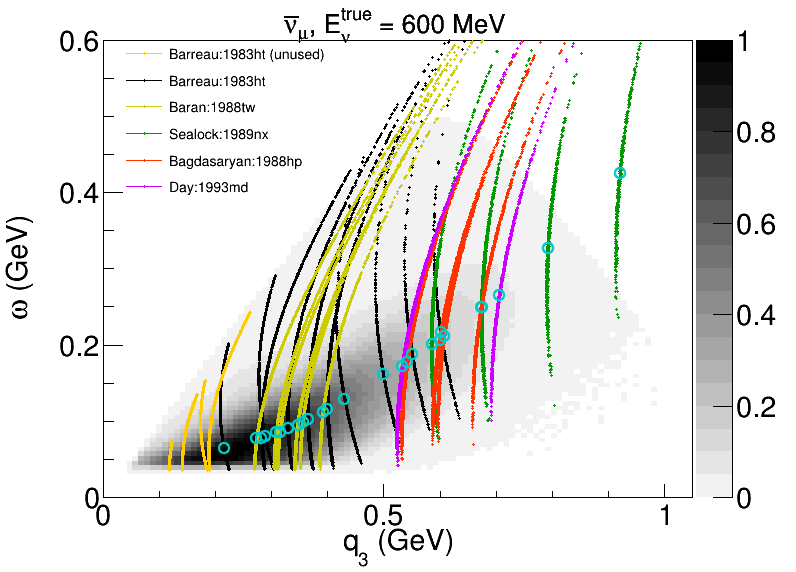}
\end{minipage}
\begin{minipage}[t]{0.32\textwidth} \centering
\includegraphics[width=1.0\textwidth]{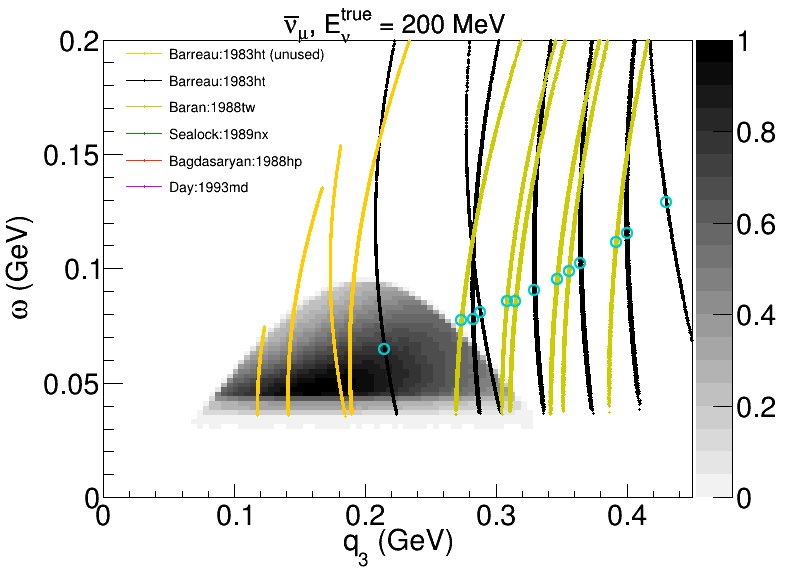}
\end{minipage}
\begin{minipage}[t]{0.32\textwidth} \centering
\includegraphics[width=1.0\textwidth]{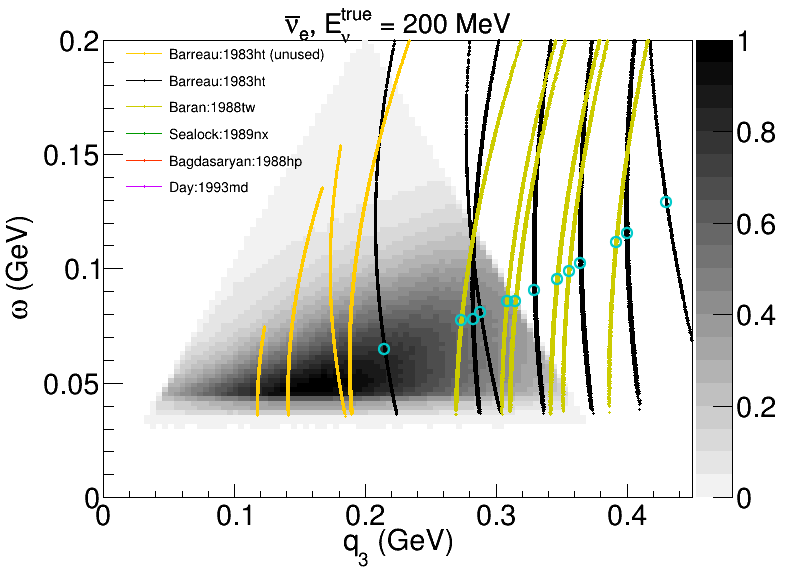}
\end{minipage}
\caption{Energy transfer $\omega$ and three-momentum transfer $q_3$ phase space of CCQE interactions and electron scattering predicted by \textsc{NEUT}.
The gray scale background shows the CCQE events on oxygen for a monochromatic neutrino energy and a given neutrino flavor, as indicated at the top of each panel.
The dots represent the electron QE events on carbon from the datasets used in this paper~\citecarbon.
The cyan open circles represent the QE peak position obtained in Sec.~\ref{sec:q3_eb}, and the orange dots represent the four datasets excluded from the analysis.
}
\label{fig:omega_q3}
\end{figure*}

To evaluate the validity and limitations of applying this correction, the coverage of energy and momentum transfer phase space is also crucial.
The electron scattering datasets used in this analysis do not fully cover the low-energy and low-momentum transfer regions, as shown in Fig.~\ref{fig:omega_q3}.
There is a small contribution in this region for high-energy neutrinos such as $E_{\nu}^{true}=600$\,MeV.
However, this contribution becomes significant for low energy or antineutrinos.
Allowing negative removal energy improves the agreement with experiments. 
However, discrepancies still remain particularly in the shape of the distribution shown in Fig.~\ref{fig:Barreau_corr_trunc}.
Thus, the correction for CCQE discussed here has a limitation, and the results should be carefully examined further.
Robust validation can be achieved in the future by adding low-energy electron scattering data to the analysis.
These datasets tend to have a fine structure, making it difficult to define peak positions.
An alternative approach for quantification will be required to fully assess the correction's accuracy.

The observed 20\,MeV shift in the $E_\nu^{rec}$ distribution is consistent with Ref.~\cite{Bodek2019}, which introduced a correction similar to the one presented in this paper.
On the other hand, the theory-based correction for the distortion effect discussed in Ref.~\cite{PhysRevD.91.033005} results in a larger shift of about 30\,MeV.
When we allow negative removal energy, we obtain a result consistent with the theory-based correction.
However, this assumption is unphysical.
The theory-based method, which maintains physical consistency and provides better agreement with experimental data, is preferable and calls for improved implementation in the future.

%%%%%%%%%%%%%%%%%%%%%%%%%%%%%
\section{Conclusion and prospects} \label{sec:conclusion}
This paper presents the successful implementation of electron scattering in the neutrino event generator \textsc{NEUT} for two interaction modes: QE and $1\pi$.
A framework is established to extend \textsc{NEUT} for electron scattering simulation, ensuring compatibility with the existing framework.
It enables direct evaluations of \textsc{NEUT} models by using abundant high-precision electron scattering data.
An SF-based model is implemented for QE interaction.
This model can describe both neutrino and electron scatterings by changing only the coupling constant and form factors.
For $1\pi$ interactions, the DCC model is employed, with code provided by the authors that describes both neutrino and electron scatterings.
\par
The \textsc{NEUT} QE implementation is validated through comparisons with numerical calculations by Ankowski {\it et al.}
The \textsc{NEUT} predictions are compared with inclusive electron scattering data of $^{12}$C$(e,e')$ and $^{16}{\rm O}(e,e')$.
A shift in the QE peak positions is observed, which strongly depends on momentum transfer.
It yields significant shifts for low momentum transfers of $q_3\lesssim0.7$\,GeV, while it is negligible for high-momentum transfer.
The correlation between the peak shift and the momentum transfer is parametrized by a linear function.
\par
The result can be used as a correction term to the removal energy, empirically introducing effects beyond the PWIA.
To evaluate its impact on neutrino oscillation experiments, the correction is applied to neutrino CCQE interactions.
We observed sizable shifts in reconstructed neutrino energy distribution of approximately 20--30\,MeV between $200$ and $1000$\,MeV for true neutrino energies.
The electron scattering data used in this study do not fully cover the energy transfer and three-momentum transfer phase space.
This issue is crucial for the validation of this correction for low-energy CCQE interactions and antineutrinos.
Further careful examination using low-energy datasets is needed.
In addition, although this paper introduces an empirical correction to the removal energy, a theory-driven correction, as discussed in Ref.~\cite{PhysRevD.91.033005}, would be preferable.
This is one possible future development for \textsc{NEUT}.
\par
This paper primarily focuses on the implementation details and discussions of QE in inclusive scattering $(e,e')$.
There are numerous exciting topics to explore in future studies.
Comparisons with recent experimental data taken at Jefferson Lab.~\cite{Khachatryan2021,PhysRevC.98.014617} and MAMI ~\cite{Mihovilovic2024} would be highly beneficial.
Particularly, comparisons with semi-inclusive measurements, such as $(e,e'p)$ and $(e,e'\pi)$, provide abundant information for validating FSI cascade models, as discussed in Ref.~\cite{Khachatryan2021}.
A deeper investigation of $1\pi$ models is also an important future task.
Currently, \textsc{NEUT} assumes that the target nucleon has an on-shell mass.
We plan to initially investigate the validity of this assumption and also to incorporate removal energy as discussed in Ref.~\cite{PhysRevC.100.045503} for further analysis.
Also, implementing other interaction channels such as multinucleon interactions and DIS will be a significant future challenge.
Completing the implementation of electron scattering will provide deeper insights into nuclear models and their connection to neutrino interactions.
\par
The code developed in this study is included in the latest version of \textsc{NEUT} 5.9.0.
Thus, further discussions on electron scattering using \textsc{NEUT} are expected to proceed smoothly.
Such discussions can be extended straightforwardly to neutrino interactions and to neutrino physics analyses using \textsc{NEUT}, e.g., analyses at Super-Kamiokande~\cite{PhysRevD.109.072014}, T2K~\cite{PhysRevD.108.072011}, and Hyper-Kamiokande~\cite{protocollaboration2018hyperkamiokandedesignreport} experiments.
The study presented here is expected to serve as a foundational step for enhancing future neutrino oscillation analyses in these experiments using electron scattering.

%%% Acknowledgments %%%
\begin{acknowledgments}
This work was supported by JSPS KAKENHI Grant No.~23KJ0319.
The author expresses deep gratitude to the host researcher of this JSPS fellowship, Professor Yoshinari Hayato, for his invaluable assistance in improving the manuscript.
The author thanks Dr. Artur M. Ankowski for his insightful comments on the manuscript and for providing his numerical calculation results, which were essential for validating the implementation.
The author also thanks Professor Toru Sato and Dr. Noemi Rocco for sharing their numerical calculations on MEC.
The author finally thanks T2K Neutrino Interaction Working Group for their valuable discussions.
\end{acknowledgments}

%%% Appenddix %%%
\appendix

%%%
\section{Event generation process of quasielastic interactions in NEUT} \label{sec:qe_neut}
In the event generation process of quasielastic interactions, the elementary cross section described in Eq.~(\ref{eq:ele_furmanski}) is transformed into the center-of-mass variables.
The velocity of the center-of-mass frame $\bold{v}$ and the Lorentz factor $\gamma$ are expressed as
\begin{equation}\begin{split}
\bold{v} & =  \frac{\bold{p}+\bold{k}}{E_k + M - \tilde{E}} = \frac{\bold{p}+\bold{k}}{y},\\
y &\equiv E_k+M-\tilde{E}, \\
\gamma & = \frac{1}{\sqrt{1-\bold{v}^2}},
\end{split} \end{equation}
where $y$ represents the total energy of the center-of-mass frame.
The Mandelstam variable $s$ of the center-of-mass frame is defined by 
\begin{equation}\begin{split}
s=y^2 - (\bold{p}+\bold{k})^2.
\end{split} \end{equation}
Since the Mandelstam variable $s$ is Lorentz invariant,
the momentum of outgoing lepton in the center-of-mass frame, $\bold{k'_{com}}$, is determined by momentum conservation as
\begin{align} \label{eq:kcom'}
|\bold{k'_{com}}| = \sqrt{ \frac{( s + m^2 - M^2)^2}{4s} -m ^2},
\end{align}
where $m$ is the outgoing lepton mass.
Finally, the total cross section in terms of the center-of-mass variables is expressed as
\begin{widetext} 
\begin{align}\label{eq:final_total_xsec}
\frac{d\sigma_{tot}}{dQ^2}
& = \frac{C}{E_k} \int d^3p\,d\tilde{E}\,P_{\text{hole}}(\bold{p},\tilde{E})
\int \frac{L_{\mu\nu}H^{\mu\nu}}{E_p E_{p'}E_{k'}}
  \frac{\sqrt{1 + (\gamma^2-1)(\cos^2\theta_{com}-1)}}{|\bold{v_{k'}}-\bold{v_{p'}}|}
  |\bold{k'}_{com}|^2 d\phi_{com}d\cos\theta_{com},
\end{align}
\end{widetext}
where $\bold{v_{k'}}$ and $\bold{v_{p'}}$ denote the velocities of outgoing lepton and nucleon in the laboratory frame, respectively,
and $\theta_{com}$ and $\phi_{com}$ represent the zenith and azimuth angles, respectively, between outgoing lepton in the center-of-mass frame and velocity of the center-of-mass frame, i.e.,
\begin{equation}\begin{split}
\bold{v_{k'}} & = \bold{k'}/E_{k'},\\
\bold{v_{p'}} & = \bold{p'}/E_{p'}, \\
\cos\theta_{com} & = \frac{\bold{v} \cdot\bold{k'_{com}}}{|\bold{v}||\bold{k'_{com}}|}.
\end{split} \end{equation}
\par
The event generation proceeds through the following steps:
\begin{itemize}
\item The removal energy $\tilde{E}$ and the momentum of the target nucleon $\bold{p}$ are sampled according to the SF.
\item The outgoing lepton energy in the center-of-mass frame, $|\bold{k'_{com}}|$, is computed using Eq.~(\ref{eq:kcom'}).
\item The angles $\theta_{com}$ and $\phi_{com}$ are sampled randomly.
\item The total cross section is calculated from Eq.~(\ref{eq:final_total_xsec}).
\item The weights are determined by comparing the precalculated maximum cross section with the cross section being currently generated. 
Events are either accepted or rejected based on their weights using a random sampling method.
\end{itemize}
The cross section shown in Fig.~\ref{fig:xsec} is calculated as the average of eventwise differential cross sections, with statistical fluctuations arising from the finite number of trials in the repeated calculation process.

\section{Checks on the implementations and behaviors of quasielastic scattering} \label{sec:appen_qe}
\subsection{Validation using numerical calculation}
\begin{figure}[htbp] \centering
\includegraphics[width=0.90\columnwidth]{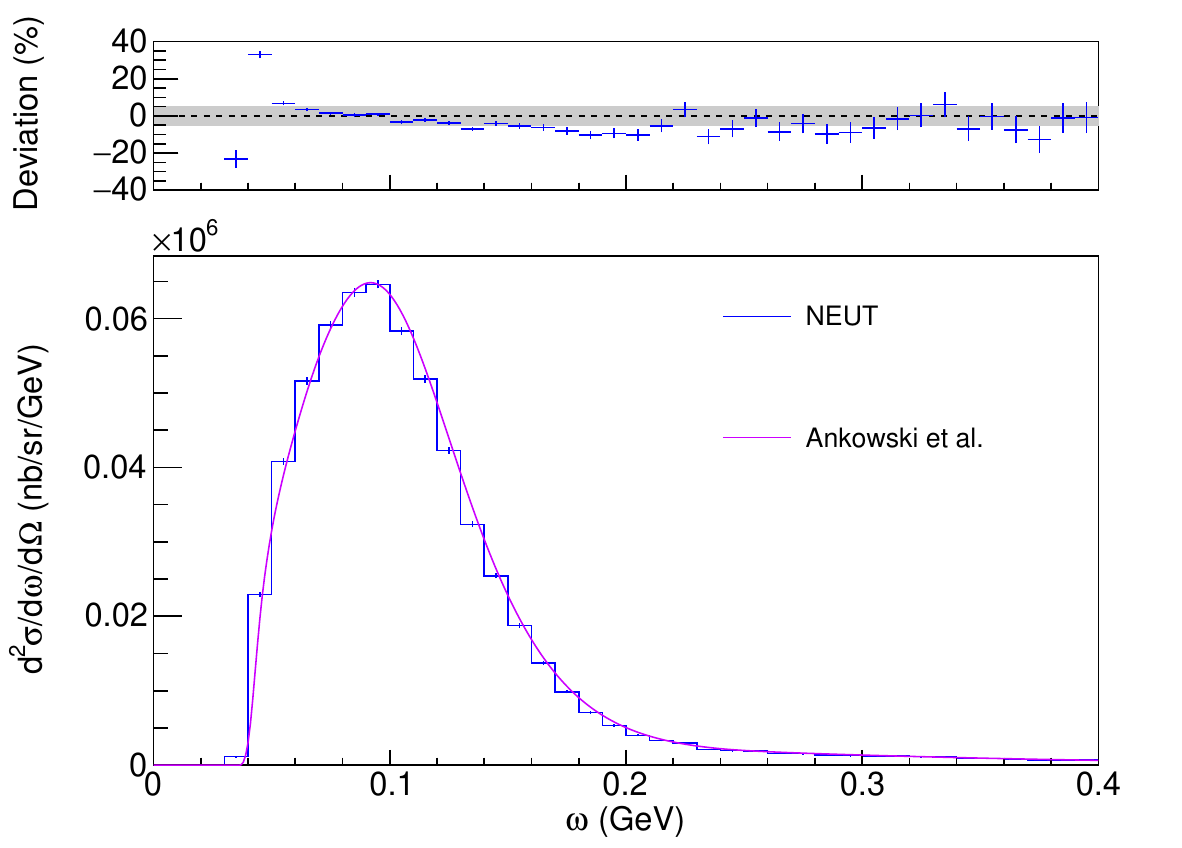}
\caption{
Comparisons of quasielastic electron scattering $^{12}$C$(e,e')$ cross section between \textsc{NEUT} (blue) and numerical calculation by Ankowski {\it et al.} (violet)~\cite{PhysRevD.91.033005}.
The electron energy is 560\,MeV and the scattering angle is $\theta=36^\circ$.
The upper panel shows the relative deviation from the numerical calculation, and the gray box represents $\pm5$\% region.
The error bar denotes the statistical error of \textsc{NEUT} simulation.
}
\label{fig:comp_neut_ankowski}
\end{figure}

The implementation of the QE model in \textsc{NEUT} is validated by comparing it with theoretical numerical calculations by Ankowski {\it et al.}~\cite{PhysRevD.91.033005}.
The \textsc{NEUT} simulation gives good agreement with the numerical calculation, with a deviation of less than $\pm5$\% around peak region, as shown in Fig.~\ref{fig:comp_neut_ankowski}.
The target nucleus is carbon, and the same Coulomb potential of 3.1\,MeV and a step-function Pauli blocking with $p_F=209\,$\,MeV are applied in these calculations.
The deviations at the rising edge of the peak are not fully understood, although they may be due to slight differences in parameters such as the masses and the finite angular acceptance of \textsc{NEUT}.
This difference is not significant to the discussion in Sec.~\ref{sec:q3_eb}, which focuses on the peak region.
Other datasets also show good consistency.
These results support the validity of this implementation.
\par
However, when these results were compared with those of McElwee {\it et al.}~\cite{psf2023008005,wreo32248}, discrepancies were observed.
These may stem from differences in analysis methods or assumptions.
Given the agreement with the numerical calculations, the findings of this study are considered valid.

%%%
\subsection{Impact of the Coulomb potential} \label{sec:appen_coul}
\begin{figure}[htbp] \centering
\includegraphics[width=0.90\columnwidth]{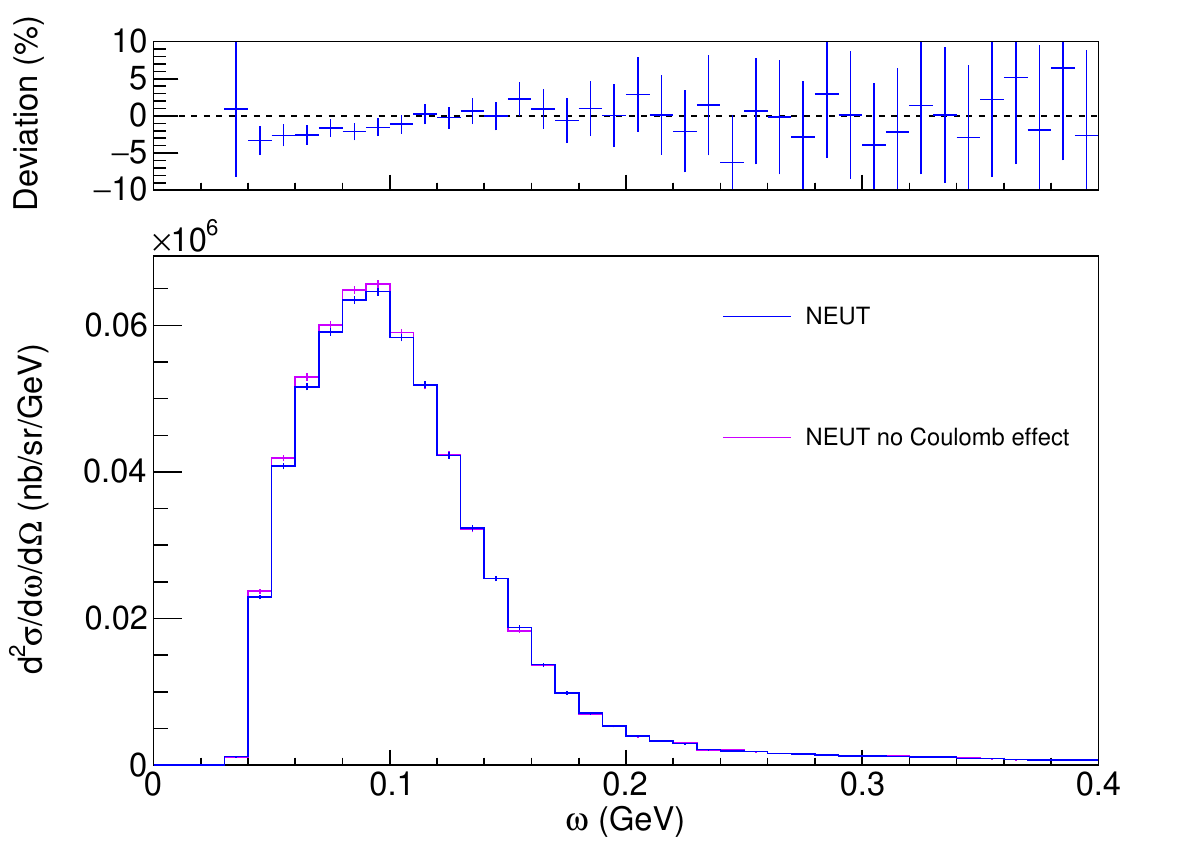}
\caption{
The same dataset as Fig.~\ref{fig:comp_neut_ankowski}, but a comparison between \textsc{NEUT} simulations with (blue) and without (violet) considering the Coulomb potential.
The upper panel shows the relative deviation from one without the Coulomb potential.
}
\label{fig:coul}
\end{figure}

The impact of Coulomb potential $|V_{eff}|$ to \textsc{NEUT} simulation is also investigated.
Figure~\ref{fig:coul} shows the cross section predicted by \textsc{NEUT} with and without the Coulomb potential.
Considering the Coulomb effect, the peak height decreases by a few percent, with a slight shift in the peak position of approximately $+0.6$\,MeV.
The impact of the Coulomb potential is small in the analysis in Sec.~\ref{sec:q3_eb}.

%%%
\subsection{Impact of Pauli blocking} \label{sec:appen_pb}
\begin{figure}[htbp] \centering
\begin{minipage}[t]{1.0\columnwidth} \centering
\includegraphics[width=1.0\textwidth]{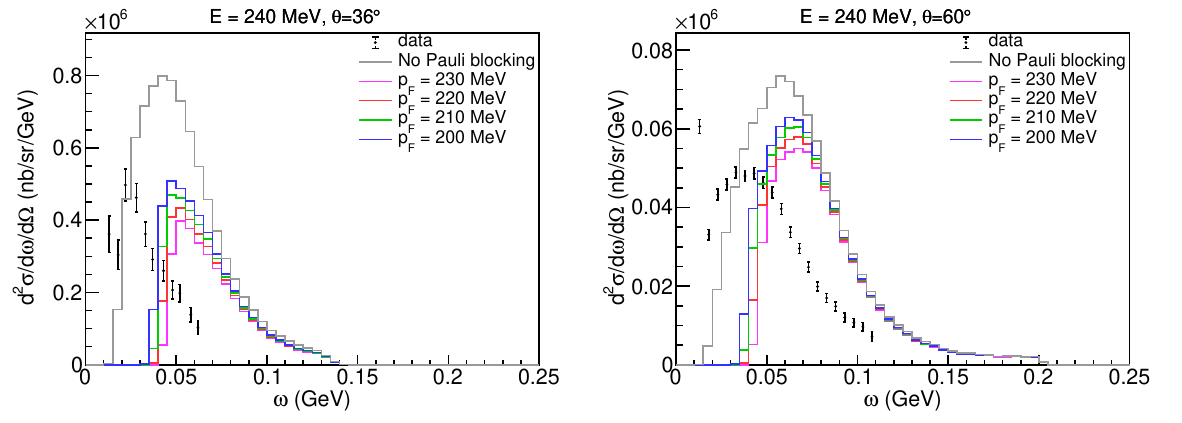}
\end{minipage}
\\
\begin{minipage}[b]{1.0\columnwidth} \centering
\includegraphics[width=1.0\textwidth]{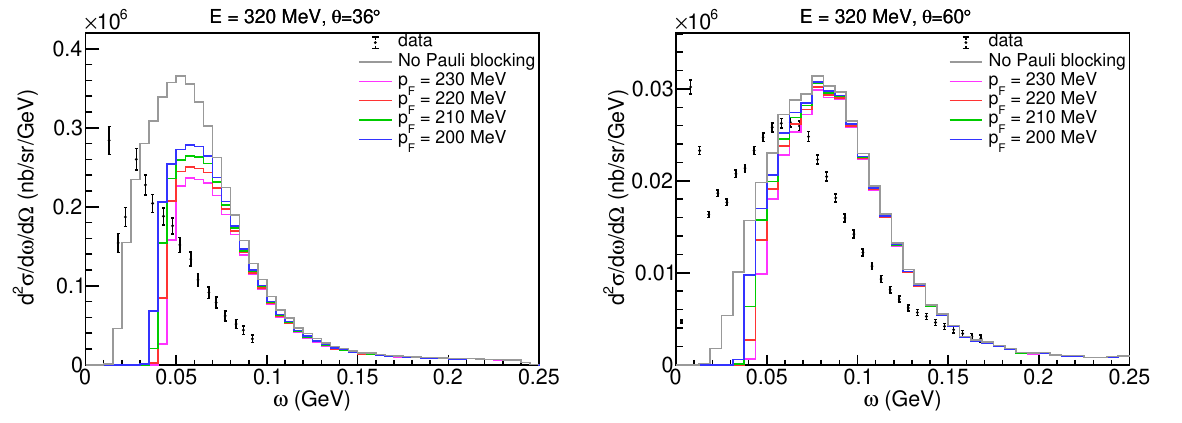}
\end{minipage}
\\
\begin{minipage}[b]{1.0\columnwidth} \centering
\includegraphics[width=1.0\textwidth]{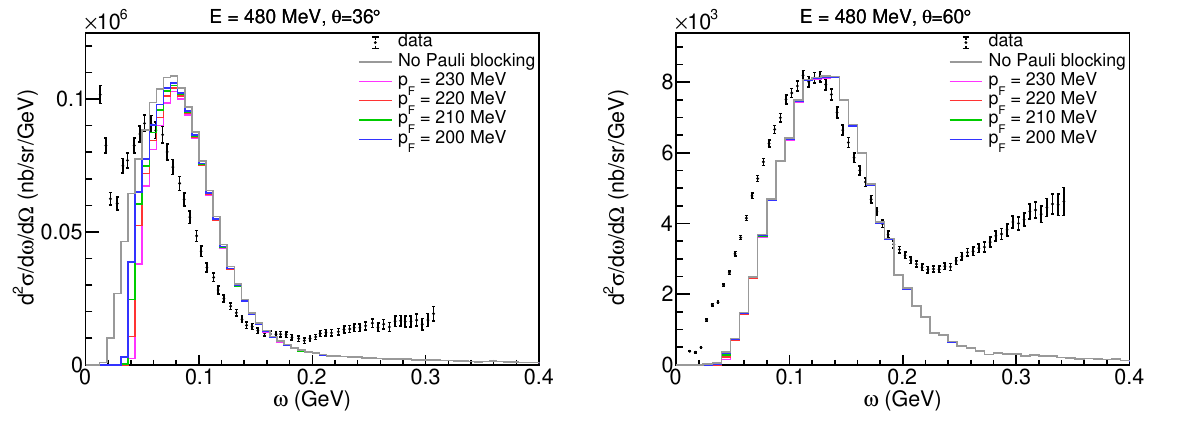}
\end{minipage}
\caption{
  Impact of Pauli blocking on $^{12}{\rm C}(e,e')$ cross section predicted by \textsc{NEUT}.
  The \textsc{NEUT} simulations consider quasielastic scattering only.
  The data is from Ref.~\cite{Barreau:1983ht}.
  The gray lines show the results without applying Pauli blocking, and the other lines show the results of varying the Fermi surface $p_F=200,\,210,\,220,\,230$\,MeV.
}
\label{fig:comp_PB}
\end{figure}

The impact of Pauli blocking is evaluated by varying the Fermi surface $p_F$ as shown in Fig.~\ref{fig:comp_PB}.
The Pauli blocking effect changes both the normalization and shape of the cross sections.
It is clear that simply varying the Fermi surface does not adequately explain the discrepancies with the data.
The impact is significant in the datasets of low-momentum transfer while becoming negligible in high-momentum transfer datasets of $Q^2\gtrsim0.1{\rm\,GeV}^2$ and $q_3\gtrsim0.3$\,GeV.
It is confirmed that varying the Fermi surface between $p_F=209$\,MeV and $p_F=\,221$\,MeV has only a minor impact on the correction function as shown in Table~\ref{tab:corr_summary}.

% WIDE
%\begin{figure*}[htbp] \centering
%\begin{minipage}[t]{1.0\columnwidth} \centering
%\includegraphics[width=1.0\textwidth]{fig_PB_dE_240.pdf}
%\end{minipage}
%\begin{minipage}[b]{1.0\columnwidth} \centering
%\includegraphics[width=1.0\textwidth]{fig_PB_dE_320.pdf}
%\end{minipage}
%\\
%\begin{minipage}[b]{1.0\columnwidth} \centering
%\includegraphics[width=1.0\textwidth]{fig_PB_dE_480.pdf}
%\end{minipage}
%\hspace{0.495\textwidth}
%\caption{
%  Impact of Pauli blocking on $^{12}{\rm C}(e,e')$ cross section predicted by \textsc{NEUT}.
%  The \textsc{NEUT} simulations consider quasielastic scattering only.
%  The data is from Ref.~\cite{Barreau:1983ht}.
%  The gray lines show the results without applying Pauli blocking, and the other lines show the results of varying the Fermi surface $p_F=200,\,210,\,220,\,230$\,MeV.
%}
%\label{fig:comp_PB}
%\end{figure*}

%%%%
%\section{Energy transfer and three-momenum transfer phase space} \label{sec:omega_q3}

%%% Bibliography %%%
\FloatBarrier
\bibliography{bib}

\end{document}